\documentclass[acmlarge,screen]{acmart}

\AtBeginDocument{%
  \providecommand\BibTeX{{%
    \normalfont B\kern-0.5em{\scshape i\kern-0.25em b}\kern-0.8em\TeX}}}

\setcopyright{acmcopyright}
\copyrightyear{2025}
\acmYear{2025}
\acmDOI{XXXXXXX.XXXXXXX}

\acmConference[Conference acronym 'XX]{Make sure to enter the correct
  conference title from your rights confirmation emai}{June 03--05,
  2025}{Woodstock, NY}
\acmBooktitle{Woodstock '25: ACM Symposium on Neural Gaze Detection,
 June 03--05, 2025, Woodstock, NY} 
\acmPrice{15.00}
\acmISBN{978-1-4503-XXXX-X/18/06}

\newcommand{\WSCoach}[0]{$\textit{WSCoach}$}
\newcommand{\Orai}[0]{$\textit{Orai}$}
\newcommand{\WSCOACH}[0]{WSCoach}
\newcommand{\ORAI}[0]{Orai}

\newcommand{\TotalUtterances}[1]{\textit{Total Utterances}}
\newcommand{\RecognitionAccuracy}[1]{\textit{Recognition Accuracy}}

\newcommand{\Filler}[1]{``#1''}

\newcommand{\RQI}[1]{\textbf{RQ1:} Does \WSCoach{} reduce the occurrence of unwanted words during the training phase{#1}?}
\newcommand{\RQII}[1]{\textbf{RQ2:} Does \WSCoach{} reduce the occurrence of unwanted words in the post-training phase{#1}?}
\newcommand{\RQIII}[1]{\textbf{RQ3:} Does \WSCoach{} outperform \Orai{} during the training phase{#1}?}
\newcommand{\RQIV}[1]{\textbf{RQ4:} Does \WSCoach{} outperform \Orai{} in the post-training phase{#1}?}

\newcommand{\RQV}[1]{\textbf{RQ5:} How does \WSCoach{} affect users' self-awareness regarding unwanted words{#1}?}
\newcommand{\RQVI}[1]{\textbf{RQ6:} How does \WSCoach{} impact the overall quality of conversation{#1}?}

\renewcommand{\quote}[1]{``\textit{#1}''}
\newcommand{\quoteby}[2]{---``\textit{#2} (#1)''}

\newcommand{\significantI}[0]{$^{\star}$}
\newcommand{\significantII}[0]{$^{\dagger}$}

\newcommand{\material}{\hyperref[appendix:meta]{META Appendix\xspace}}

\usepackage{lipsum}
\usepackage{nicematrix}
\usepackage{graphicx}  
\usepackage{multirow}
\usepackage{float}
\usepackage{booktabs}
\usepackage{verbatim}
\usepackage{enumitem}
\usepackage{xspace}
\begin{document}

\title[\WSCOACH{}]{\WSCOACH{}: Wearable Real-time Auditory Feedback for Reducing Unwanted Words in Daily Communication}


\author{Zhang Youpeng}
\email{yzhan63@cityu.edu.hk}
\orcid{0009-0003-1142-7296}
\affiliation{%
\department{Synteraction Lab}
\institution{School of Creative Media, City University of Hong Kong}
\city{Hong Kong}
  \country{China}
}

\author{Nuwan Janaka}
\email{nuwanj@u.nus.edu}
\orcid{0000-0003-2983-6808}

\affiliation{%
  \institution{Smart Systems Institute, National University of Singapore}
  \department{Synteraction Lab}
  \country{Singapore}
}

\author{Ashwin Ram}
\email{ashwinram10@gmail.com}
\orcid{0000-0003-1430-8770}
\affiliation{%
  \institution{HCl Lab, Saarland Informatics Campus, Saarland University}
  \city{Saarland}
  \country{Germany}
}

\author{Yin Peilin}
\email{}
\orcid{}
\affiliation{%
  \institution{Zhejiang University}
  \city{Hangzhou}
  \country{China}
}

\author{Tian Yang}
\email{}
\orcid{}
\affiliation{%
  \institution{Guangxi University}
  \city{Nanning}
  \country{China}
}

\author{Shengdong Zhao}
\authornote{Corresponding Author.}
\email{shengdong.zhao@cityu.edu.hk}
\orcid{0000-0001-7971-3107}

\affiliation{%
\department{Synteraction Lab}
\institution{School of Creative Media \& Department of Computer Science, City University of Hong Kong}
\city{Hong Kong}
  \country{China}
}

\author{Pierre Dragicevic}
\authornotemark[1]
\email{pierre.dragicevic@inria.fr}
\orcid{0000-0002-1854-5899}
\affiliation{%
  \institution{INRIA}
  \city{Bordeaux}
  \country{France}
}

\renewcommand{\shortauthors}{Youpeng et al.}



\begin{abstract}

The rise of wearable smart devices raises unprecedented opportunities for self-improvement through ubiquitous behavior tracking and guidance. However, the design of effective wearable behavior intervention systems remains relatively unexplored. To address this gap, we conducted controlled studies focusing on the reduction of unwanted words (e.g., filler words, swear words) in daily communication through auditory feedback using wearable technology. We started with a design space exploration, considering various factors such as the type, duration, and timing of the auditory feedback. Then, we conducted pilot studies to reduce the space of design choices and prototyped a system called \WSCoach{} (Wearable Speech Coach), which informs users when they utter unwanted words in near-real-time. To evaluate \WSCoach{}, we compared it with a state-of-the-art mobile application supporting post-hoc conversation analysis. Both approaches were effective in reducing the occurrence of unwanted words, but \WSCoach{} appears to be more effective in the long run. Finally, we discuss guidelines for the design of wearable audio-based behavior monitoring and intervention systems and highlight the potential of wearable technology for facilitating behavior correction and improvement. For supplementary material, please see the \material{} and our OSF project at \url{https://osf.io/6vhwn/?view_only=489498d3ac2d4703a17475fc6ca65dfa}.

\end{abstract}



\begin{teaserfigure}
  \includegraphics[width=\textwidth]{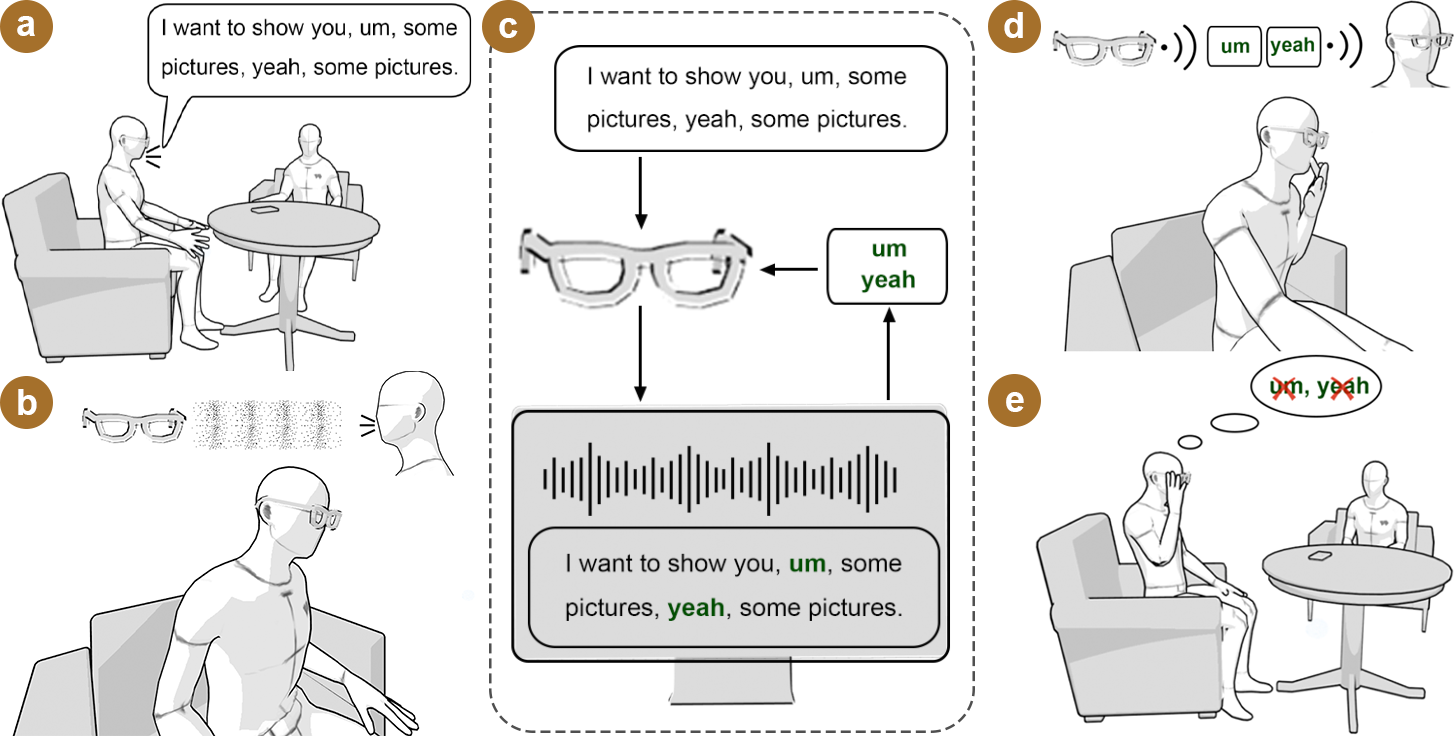}
  \caption{\WSCoach{} helps users reduce the use of words they wish not to say in daily conversations, such as filler words and swear words. (a) The user wears smart audio-based glasses and engages in daily conversations; (b) The (audio) smart glasses continuously monitors their speech, which is (c) analyzed to identify the unwanted words the user has specified in advance; (d) \WSCoach{} provides real-time auditory feedback through the smart glasses. (e) This active self-monitoring helps the user refrain from using unwanted words during subsequent interactions. The above steps also represent the experimental task in Section~\ref{sec:eval}. }
  \Description{.}
  \label{fig:study2:setting}
\end{teaserfigure}


\maketitle

\section{Introduction}

The rise of smart wearable devices has opened up unprecedented opportunities for improving our daily lives through ubiquitous behavior tracking and guidance. One promising area of research involves utilizing wearable devices to monitor users' status and deliver intelligent interventions to improve well-being and behaviors \cite{cajita_feasible_2020, scott_behavior_2024}. While visual feedback-focused wearable solutions have demonstrated effectiveness (e.g., \cite{smith_integrating_2020, tanveer_rhema_2015, damian_augmenting_2015}), it is essential to recognize that our visual channels are frequently occupied \cite{brewster_using_1997}. Consequently, audio-based real-time monitoring and intervention solutions hold considerable promise \cite{brewster_using_1997, lin_interactive_2013} despite being relatively less explored.

In particular, we aim to leverage audio-based solutions to tackle speech-related behaviors. Effective communication is vital in both personal and professional contexts \cite{seals2022we, zhao2017semi}, allowing us to navigate diversity, foster trust and respect, and cultivate an environment conducive to sharing ideas and problem-solving. Nevertheless, it is uncommon to find individuals who can entirely refrain from using filler words or repetitive expressions, especially when hurried or unprepared in speech. These unwanted words, encompassing filler words, unprofessional slang, offensive language and repetitive expressions, are widely recognized as superfluous language that can undermine the effectiveness of communication, impacting the speaker's credibility and the audience's comprehension \cite{seals2022we}.

Unlike activities like running or cycling, where existing audio-based intervention solutions have been designed and effectively used (e.g., \cite{bood2013power}), everyday conversations involve social situations where the potential interference \cite{mcatamney_examination_2006} of the feedback provided by the wearable solution to one's social acceptance needs to be carefully considered. Disruptive intervention techniques can be seen as impolite and undesirable \cite{alallah_performer_2018}, thus demanding a separate investigation. While existing interventions leverage mobile applications to assist individuals in curbing the usage of unwanted words (e.g., \cite{Orai}), these solutions may disrupt natural conversations and prove inconvenient in hands-free scenarios. 

To address this challenge, we introduce Wearable Speech Coach (\WSCoach{}), a system designed to enhance awareness through carefully designed auditory feedback, providing near real-time interventions (with feedback provided within 1 to 2 seconds) to decrease the usage of unwanted words. \WSCoach{} can be adapted to any wearable device with audio input and output, such as Bluetooth earphones or headsets. 
While wrist-worn devices and smart rings offer discreetness, they lack the ability to deliver detailed, real-time auditory feedback during conversation. Smart glasses with visual displays support richer feedback but can disrupt eye contact and visual attention—both critical for face-to-face interaction. To balance functionality and social appropriateness, we selected an audio-based smart glasses platform. This form factor enables subtle, always-available feedback with minimum interference to conversational norms. Unlike mobile phones or earphones, which may appear distracting or impolite, audio-based smart glasses provide a more seamless and socially acceptable alternative. The feedback is mostly audible only to the wearer, minimizing disruption to others and supporting natural, real-time interaction. Additionally, their hands-free, heads-up design supports mobility, multitasking, and privacy—aligning with our goal of delivering discreet, context-sensitive support. These qualities also reflect the heads-up computing paradigm \cite{zhao_heads_up_2023}, which seeks to embed computational support seamlessly into everyday social interactions. While smart glasses remain an emerging platform, their expanding ecosystem and growing market presence suggest strong potential for broader adoption.

We conducted a series of pilot studies to identify the optimal attributes of auditory feedback (i.e., type, duration, timing), ultimately revealing that spearcon emerged as the most preferred feedback mechanism based on consistently higher positive ratings across various metrics. Building on these findings, we evaluated the efficacy of \WSCoach{} in reducing the occurrence of unwanted words during conversations, comparing its performance to the professional mobile application, ``\Orai{}''. The results confirmed the effectiveness of \WSCoach{} in decreasing unwanted words.

The contributions of this paper are thus threefold:
1) Understand and empirically evaluate the suitability of various auditory feedback to improve awareness of speaking unwanted words during daily communications.
2) An artifact that helps people reduce unwanted words using wearable technology with real-time auditory feedback.
3) An empirical study that evaluates the effectiveness of decreasing unwanted words with a wearable system, offering further design implications.

\section{Related Work} 
\label{sec:related_work}

Speech disfluencies, such as filled pauses (e.g., “uh”, “um”), tongue clicks, and frequently used fillers (e.g., “like”, “you know”), are prevalent in everyday communication and can hinder effective communication \cite{mancuso_using_2016}. People are motivated to minimize such behaviors (e.g., unwanted words) and improve their daily communication \cite{menjot_interventions_2023}. Recent advances in wearable technologies have opened new avenues for interventions aimed at enhancing communication \cite{Bubel2016Awareme, damian_augmenting_2015, muralidhar2016dites}. Our studies focus on reducing unwanted words in conversations using (near) real-time speech intervention with auditory feedback, contributing to the broader objective of facilitating real-time speech improvement through wearable devices. Thus, our research is related to:

\subsection{Speech Interventions and Awareness Training}
\label{sec:related_work:speech_intervention}

Speech interventions aim to enhance communication proficiency and reduce speech disfluencies \cite{montes_awareness_2019, ortiz2022decreasing, spieler_using_2017, hazel_immediate_2011}. While there are different types of speech interventions (see \cite{menjot_interventions_2023} for a review), awareness training is a common and effective strategy to reduce speech disfluencies \cite{seals2022we, menjot_interventions_2023, himle2006brief}. Awareness training involves the conscious identification of specific undesirable speech patterns, such as filled pauses or the overuse of filler phrases like “like” and “you know,” facilitated by various means, including audio and video recordings \cite{menjot_interventions_2023}.

Awareness training ranges from (near) real-time methods\footnote{Note: There is a delay between actual occurrence and alerting feedback due to the detection time.}, where a trainer signals the occurrence of disfluencies (e.g., by alerting feedback) \cite{spieler_using_2017, hazel_immediate_2011, Siegel1967Verbal}, to post-training analysis (e.g., retrospective feedback) through audio/video recordings to pinpoint disfluencies \cite{ortiz2022decreasing, cavanagh_effect_2014}. Early real-time methods suffered from a generic feedback approach, which lacked specificity for different types of disfluencies. For example, these early works used a human observer monitoring the user’s speech behaviors and raising hands \cite{montes_awareness_2019, spieler_using_2017} or initiating a single auditory alert \cite{hazel_immediate_2011, Siegel1967Verbal} upon detecting any speech disfluency. While this approach enabled users to identify the disfluencies, it prevented users from distinguishing between them to have more detailed awareness (e.g., separating the filler word “like” from “you know”). 

Post-training techniques mitigate this lack of specificity by analyzing recorded speech to detect disfluencies and offering targeted suggestions as retrospective feedback, but they demand additional time and effort for post-analysis. Most of these post-analysis methods use desktop or mobile platforms to provide detailed feedback due to their processing and feedback capabilities. For example, VoiceCoach \cite{wang_voicecoach_2020} is an interactive desktop application targeting public speaking that analyzes voice modulation skills, recommends suitable examples (e.g., from TED talks), and provides visual feedback on performance (e.g., pause, volume, pitch, speed). Orai \cite{Orai} is an interactive mobile application that provides retrospective feedback on speech components such as filler words, pace, and conciseness. It outputs a detailed analysis of recorded speeches via annotated transcripts, highlights areas of improvement in the user’s speech, and helps keep track of progress. ELSA Speak \cite{corporation_elsaspeak_2024} improves English speaking, specifically pronunciation, by utilizing AI to identify errors and provide visual feedback on recorded audio using a mobile application. StammerApp \cite{mcnaney_stammerapp_2018} is a mobile application designed to support people who stammer. It allows users to set goals related to challenging real-world speaking situations with recorded video/audio (e.g., ordering food, booking a taxi) and practice to overcome side effects (e.g., word repetition, regular use of interjections) during those situations .

Considering the trade-offs in previous systems---early (near) real-time feedback lacking specificity and post-intervention methods having specificity but requiring time or motivation to reflect---we consider the potential of (near) real-time feedback to support awareness training while maintaining specificity to minimize speech disfluencies.

\subsection{Wearable Real-Time Interventions and Feedback Modalities}
\label{sec:related_work:feedback_modality}

Wearable technology enables us to receive suitable real-time assistance (e.g., reminders from smartwatches, language translations from earbuds, navigation guidance from smart glasses) anywhere with little user effort. Wearable (near) real-time interventions have been commonly used in awareness training, including speech training \cite{damian_augmenting_2015, tanveer_rhema_2015, Bubel2016Awareme, mihoub_social_2017}.

Most of these interventions use visual feedback due to its increased modality capacity \cite{rau_modality_2019} to show specific details. For example, Rhema \cite{tanveer_rhema_2015} uses smart glasses to provide real-time visual feedback on the speaker's volume (e.g., make louder) and speech rate (e.g., slow down) during public speaking. Logue \cite{damian_augmenting_2015} analyzes verbal and non-verbal behavior during public speaking using sensors and provides visual feedback on behavior (speech rate, energy, openness, appropriateness) on smart glasses in an unobtrusive way. However, in conversations unlike public speaking, such visual feedback can disrupt social interactions due to reduced eye contact when attending to visual feedback \cite{mcatamney_examination_2006, janaka_paracentral_2022}.

Thus, alternative feedback modalities like haptic feedback have been explored. AwareMe \cite{Bubel2016Awareme} is a wristband that provides real-time haptic feedback for anxiety detection during presentations. While haptic feedback is subtle and less interfering \cite{tam_design_2013}, it has limited modality capacity to show detailed feedback \cite{rau_modality_2019, Bubel2016Awareme}. As a result, AwareMe \cite{Bubel2016Awareme} also used visual feedback to indicate pitch, speed, and filler words.

Thus, auditory feedback emerges as an alternative that strikes a balance between these two factors as it has higher modality capacity \cite{rau_modality_2019}. Therefore, real-time auditory feedback alone has been explored in various behavior interventions \cite{vorbeck2020using, krukauskas_using_2019, quinn_evaluation_2017}. Radhakrishnan et al. \cite{radhakrishnan2021applying} utilized wearables for real-time auditory reminders against unhealthy head postures, Casamassima et al. \cite{casamassima_wearable_2014} for gait correction, and Md Mahbubur et al. \cite{rahman_breathebuddy_2022} for guided breathing exercises.

\subsection{Potential of Real-Time Auditory Feedback-Based Speech Interventions}

In addition to the above-mentioned advantages, wearable auditory feedback systems can be more energy efficient than their visual feedback counterparts and thus have longer usage durations (e.g., auditory smart glasses generally last longer than visual smart glasses on a single charge \cite{re_ak_introduction_2024}).

Existing (near) real-time auditory feedback systems in speech training typically target a general group of disfluencies (e.g., filler occurrences) \cite{hazel_immediate_2011} without providing detailed feedback (e.g., specific filler words). Given the higher modality capacity \cite{rau_modality_2019}, (near) real-time auditory feedback can be used to indicate specific target behaviors during awareness training, which is underexplored in speech interventions. This underutilization may be due to the fact that auditory feedback during speech may disrupt natural speech flow by interfering with the speaking process itself (e.g., auditory masking\footnote{Masking refers to the reduction in sound clarity due to interference from another sound, such as the decreased comprehensibility of speech overlaid with background noise \cite{peres_chapter_2008}.}) \cite{peres_chapter_2008, jacks_auditory_2015}. 

Furthermore, whether providing detailed auditory feedback in speech interventions can improve awareness and which types of auditory feedback are more suitable lacks systematic exploration in the literature. Thus, our work explores suitable (near) real-time auditory feedback for awareness training, utilizing wearable technology to develop and evaluate auditory feedback for speech correction, particularly in reducing unwanted words during conversations. Such interventions will highlight the potential of wearable intelligent assistants using auditory feedback and heads-up computing \cite{zhao_heads_up_2023, zhou2024glassmail, felicia2023mindful}.

\section{Literature Review: Auditory Feedback for Speech Interventions}

To outline a design space of auditory feedback suitable for speech intervention, we reviewed papers from the fields of HCI and Speech Communication. Our review process focused on two guiding questions: 1) What constitutes auditory feedback interventions? and 2) What measures are suitable for evaluating the effectiveness of auditory feedback for speech interventions?

\subsection{Methodology}

To construct this design space, we employed a design space analysis method \cite{maclean_questions_1991, elliott_living_2017}. Initially, we conducted searches in the ACM Digital Library, IEEE Xplore, Scopus, and Google Scholar using related keywords such as \quote{auditory feedback}, \quote{speech improvement}, \quote{speech intervention}, \quote{filler words}, \quote{self-awareness}, and \quote{communication aids}, following related systematic reviews \cite{nees_auditory_2023, frauenberger_survey_2007, menjot_interventions_2023, pinder_digital_2018}. Our inclusion criteria comprised peer-reviewed publications aligned with our objectives, excluding works focused on non-auditory feedback or interventions, non-English publications, and publications before 2009. Subsequently, we reviewed the references of these relevant publications and included seminal works that introduced types of auditory feedback and auditory interventions, resulting in a total of 27 papers. Finally, we manually reviewed these papers and extracted themes that matched our objective (e.g., auditory feedback and its suitability for speech intervention).

\subsection{Design Space}
\label{sec:design_space}

\begin{figure*}[hptb]
	\centering
    \includegraphics[width=\textwidth]{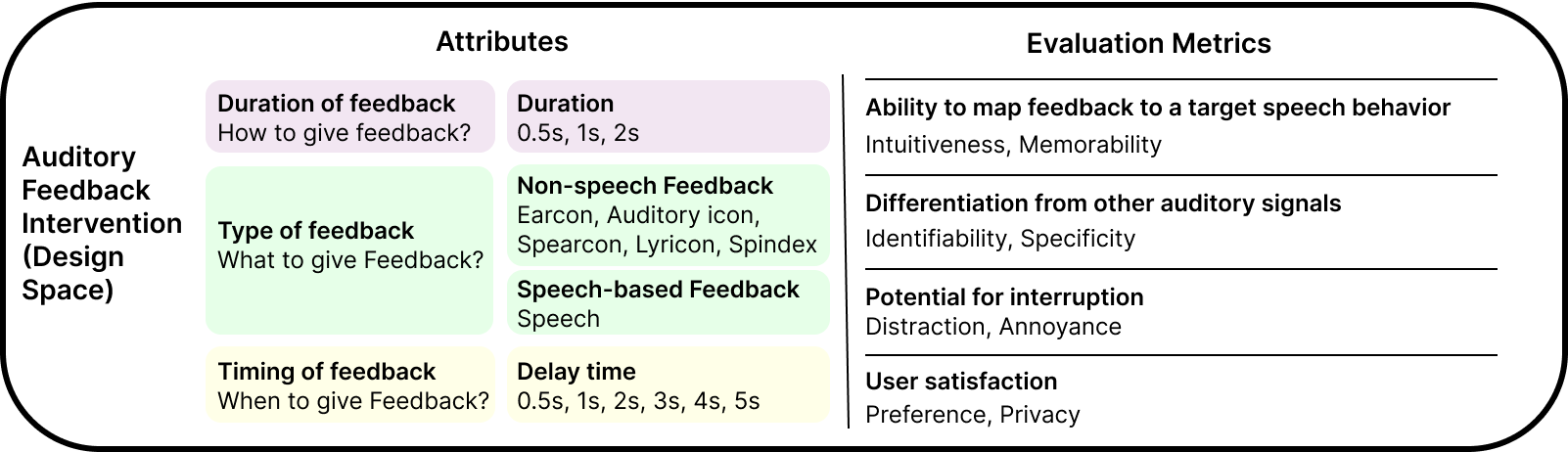}
	\caption{The design space of auditory feedback for speech interventions.}
    \label{fig:literature_review:design_space}
\end{figure*}

Based on our literature review, we have created a design space, as illustrated in Figure~\ref{fig:literature_review:design_space}, which encompasses attributes of auditory feedback (e.g., how to feedback, what, when) and evaluation criteria for speech interventions.

\subsubsection{Auditory Feedback Intervention}
\label{sec:design_space:auditory_feedback}

The design space of auditory feedback intervention comprises three attributes: how to give feedback, what to give as feedback, and when to give feedback. 
While we note that there are granular-level generic auditory feedback properties (e.g., frequency, loudness \cite{peres_chapter_2008, cabral_auditory_2019, jeon_vehicle_2022}) that can affect speech interventions, we do not consider them within our design space as they are determined when an auditory signal is designed (e.g., frequency/pitch) or are easily adjustable (e.g., loudness/amplitude).

\paragraph{Duration of Feedback: How to give feedback?}
Feedback duration is a critical factor that affects the user's experience during interventions \cite{cabral_auditory_2019}.
For example, a prolonged sound can be distracting, especially if it interrupts ongoing tasks or occurs in a context where sustained attention is required. Conversely, a too-short sound can be difficult to identify, leading to a failure to respond appropriately to the feedback. While there are some guidelines on auditory feedback duration---such as more than 200 ms for vehicle warnings \cite{lewis_validation_2018, jeon_vehicle_2022}, and between 400 ms and 2000 ms for auditory icons \cite{cabral_auditory_2019}---there is no consensus on the appropriate duration for speech interventions.

\paragraph{Type of Feedback: What to give as Feedback?}
Auditory feedback can be divided into two broad cue types: non-speech and speech-based feedback based on their common usage in digital interfaces \cite{nees_auditory_2023, jeon_vehicle_2022, peres_chapter_2008, csapo_overview_2013} (Please refer to the video figure to hear examples of each). 
Non-speech feedback includes earcons \cite{blattner_earcons_1989}, which are non-speech audio cues, typically consisting of short, rhythmic sequences of musical notes with variable intensity, timbre, and register; auditory icons \cite{gaver_auditory_1986, mynatt_designing_1994, cabral_auditory_2019}, which are nonmusical sounds that resemble (either literally or metaphorically) the objects, functions, and events they represent; spearcons \cite{walker_spearcons_2013}, which are brief sounds produced by speeding up spoken phrases, sometimes to the point where the resulting sound is no longer comprehensible as speech; lyricons \cite{jeon_lyricons_2013}, which combine the lyrics (i.e., speech) with earcons; and spindexes \cite{jeon_spindex_2011}, which are created by associating an auditory cue with an item (e.g., menu item), where the cue is based on the pronunciation of the first letter of each item. 
In contrast, speech-based feedback consists of spoken verbal cues or phrases. 
Given the advantages and disadvantages of cue type in various application scenarios \cite{nees_auditory_2023, peres_chapter_2008, garzonis_auditory_2009}—such as auditory icons that rely on familiar, real-world sounds for intuitive understanding but might need learning to associate with specific actions; earcons that are flexible, abstract sounds suitable for hierarchical information but challenging to learn without inherent meaning; and spearcons that combine the clarity of speech with the flexibility of earcons and are less recognizable as speech, potentially reducing interference but may still suffer from speech-related issues like noise interference—the type of auditory feedback significantly influences its suitability for speech interventions.

\paragraph{Timing of Feedback: When to give Feedback?} 
Auditory feedback can be used to indicate target behavior (e.g., unwanted words) during speech interventions (e.g., awareness training \cite{menjot_interventions_2023}) and reduce such behaviors \cite{montes_awareness_2019, hazel_immediate_2011, mancuso_using_2016, menjot_interventions_2023}. However, due to the temporal nature of the auditory modality and limitations of auditory sensory memory (a.k.a., phonological short-term memory or echoic memory) \cite{cowan2008differences, darwin1972auditory}, such feedback can be provided either immediately or with a slight delay of up to 4 seconds to maintain in short-term memory for identification \cite{darwin1972auditory}. Moreover, the optimal timing for feedback may depend on the situation. For example, in high-stakes or safety-critical environments, immediate feedback might be crucial \cite{lewis_validation_2018}. In contrast, in less critical learning situations, a brief delay might allow for reflection before receiving feedback \cite{hazel_immediate_2011}. Therefore, the `delay time', the temporal gap between the target behavior (e.g., the utterance of unwanted words) and the corresponding auditory feedback, can influence the effectiveness of the auditory feedback intervention.

\paragraph{Summary} 
In auditory feedback for speech interventions, selecting between non-speech and speech-based cues requires careful consideration of the context-specific advantages.  Feedback duration is a critical factor that necessitates a balance between impact and the potential for distraction. Timing is crucial, as the strategic delivery of feedback during speech hinges on the temporal nature of auditory perception and the memory constraints of users. While certain contexts provide established guidelines (e.g., vehicle warnings \cite{lewis_validation_2018, jeon_vehicle_2022}), due to the lack of such guidelines for speech interventions, identifying the optimal feedback attributes for speech interventions is a key area for further research. This underscores the intricate interplay between duration, type, and timing of auditory feedback in enhancing communication effectiveness.

\subsubsection{Evaluating Auditory Feedback Intervention}
\label{sec:design_space:metrics}

Our literature review revealed a range of metrics for evaluating auditory feedback interventions across various applications \cite{garzonis_auditory_2009, jeon_vehicle_2022, peres_chapter_2008, mynatt_designing_1994, stanton_influence_2020, jeon_spindex_2011, jeon_lyricons_2013, walker_spearcons_2013}. We compiled a list of 20 metrics to identify the most relevant metrics for speech interventions and engaged three co-authors in an independent review process. This collaborative effort led to the consensus selection of eight key metrics.

While considering these metrics, we assessed their applicability to our specific context. For example, the International Standards for Auditory Display Guidelines in Vehicles (ISO 15006) recommend metrics such as Audibility, Comprehensibility, Distinguishability, and Safety criticality \cite{jeon_vehicle_2022, 1400_1700_iso_2011}. Given that our intervention scenarios primarily involve stationary settings rather than dynamic environments like driving, we deemed the Safety criticality metric less relevant and excluded it from our selection.

These metrics are essential for understanding the alignment between feedback objectives and user expectations. They include the ability to map feedback to a target speech behavior (e.g., Intuitiveness, Memorability), differentiation from other auditory signals (e.g., Identifiability, Specificity), the potential for interruption (e.g., Distraction, Annoyance), and user satisfaction (e.g., Preference, Privacy). 
\begin{itemize}[leftmargin=*]
    \item \textbf{Intuitiveness} \cite{garzonis_auditory_2009, cabral_auditory_2019, jeon_vehicle_2022, mynatt_designing_1994} assesses the user's ability to understand the feedback's connection to the target speech behavior. 
    \item \textbf{Memorability} \cite{garzonis_auditory_2009, nahar2022interactive} evaluates how easily users can remember the meaning associated with specific auditory feedback. 
    \item \textbf{Identifiability} \cite{cabral_auditory_2019, jeon_vehicle_2022} determines whether the auditory feedback is easily distinguishable from other environmental sounds. 
    \item \textbf{Specificity} \cite{jeon_vehicle_2022, ashby_king_defining_2021} measures the precision with which the feedback identifies the target speech behavior. 
    \item \textbf{Distraction} \cite{muralidhar2016dites, francombe2013modelling} examines the degree to which the feedback might interrupt the speech flow. 
    \item \textbf{Annoyance} \cite{jeon_spindex_2011, montes_awareness_2019} evaluates any discomfort or irritation caused by the feedback. 
    \item \textbf{Preference} \cite{garzonis_auditory_2009, cabral_auditory_2019, jeon_spindex_2011} investigates the types of feedback users find most agreeable and useful.
    \item \textbf{Privacy} \cite{patil2015interrupt, kosch2016comparing} considers the likelihood of the feedback inadvertently revealing sensitive information.
\end{itemize}

\section{Pilot Studies: Designing Auditory Feedback to Minimize Unwanted Word Usage}

To address our research question, \textit{``What are the suitable attributes (i.e., duration, type, delay time) of auditory feedback for reminding users of unwanted word usage during daily conversations?''}, we conducted three iterative pilot studies. Each study focused on one attribute. 
All user studies (the pilot studies here and the evaluation study we will report in Sec~\ref{sec:eval}) have been approved by the Institutional Review Board at our university.

\subsubsection{Participants}
We recruited six participants (P1--P6, 4 male, 2 female, mean age = 23, SD = 1.4), all with no reported auditory impairments and self-identified as fluent English speakers, from the university community. All six participants did all three pilot studies, which took place on different days.

\subsubsection{Apparatus}
During conversations, participants wore smart audio glasses (Huawei Eyewear version 1\footnote{Web page from the manufacturer archived at \url{https://web.archive.org/web/20221007024433/https://consumer.huawei.com/en/wearables/huawei-eyewear/}. The Huawei Eyewear is crafted for continuous wear (11 hours of audio playback), featuring discreet/subtle acoustic outlets. The product has been replaced by a version 2.} with lenses removed, 31g, stereo audio) connected to a mobile phone (Huawei P10) via Bluetooth to receive auditory feedback on unwanted word usage. The mobile phone was equipped with a custom-built Android application capable of triggering various types of auditory feedback, each with different durations and delay times.

\subsection{Task and Procedure}

\begin{figure*}[hptb]
	\centering
	\includegraphics[width=1\linewidth]{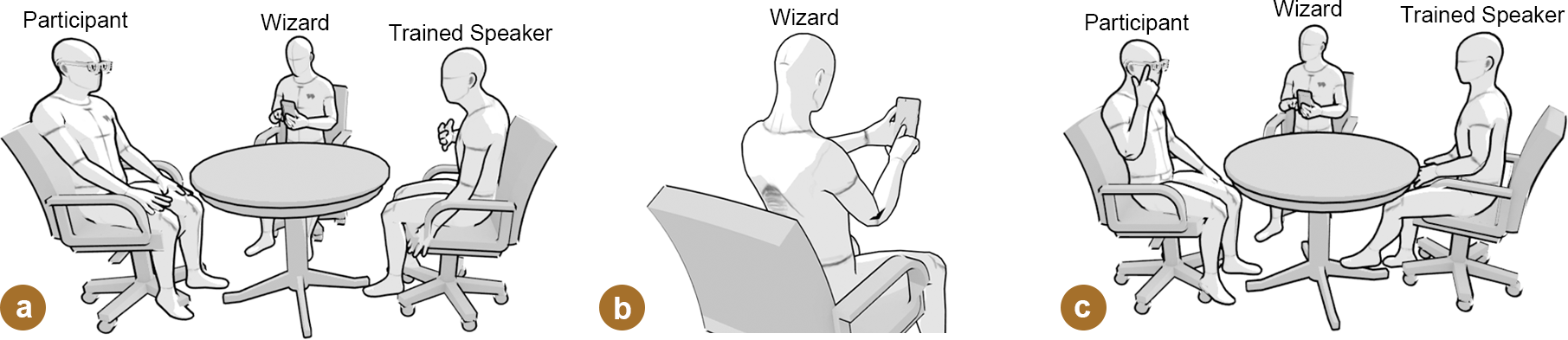}
	\vspace*{-1mm}
	\caption{Procedure for the pilot studies: (a) A participant, seated with two experimenters, engages in a conversation while wearing smart audio glasses. One experimenter (the trained speaker) converses with the participant, while the other (the wizard) operates the mobile app. (b) Upon detecting unwanted words spoken by the participant, the wizard activates auditory feedback via the app. (c) The participant receives auditory feedback through the smart audio glasses.}
	\vspace*{-1mm}
	\label{fig:WozStudy}
\end{figure*}

All pilot studies used Wizard of Oz testing (\autoref{fig:WozStudy}). After obtaining consent and being provided with a briefing, each participant was asked to list unwanted words they frequently use in conversations and wished to minimize, encompassing both non-speech (e.g., \Filler{Um}, \Filler{Uh}) and speech (e.g., \Filler{I mean}, \Filler{Like}) words. 
Before the start of each pilot study, each participant engaged in a 10-minute training session encompassing every condition to get familiar with the feedback. During these sessions, they received auditory feedback designed for each condition, receiving feedback upon speaking the unwanted words they aimed to reduce. A trained speaker (acting as the conversation partner) facilitated the conversations, encouraging participants to speak on communication topics derived from IELTS Speaking Questions (\url{https://ielts.org/}). Another trained experimenter, referred to as the 'wizard,' monitored the conversation and activated predetermined auditory feedback whenever an unwanted word was detected.
This auditory feedback was delivered to the participant through smart audio glasses, reminding them of their unwanted word usage, in line with awareness training practices \cite{menjot_interventions_2023}. After each session, the participant rated each condition of auditory feedback based on specified metrics (e.g., Distraction, Preference, Sec~\ref{sec:design_space:metrics}) relevant to each design, using 7-point Likert scales.

\subsection{Study Design}

We conducted three pilot studies (Pilot 1: identifying suitable duration, Pilot 2: identifying suitable feedback type, and Pilot 3: identifying the acceptable feedback delay, in sequential order) to examine different \textbf{design attributes of auditory feedback}. Each study's findings informed the subsequent study, ensuring an iterative refinement process. 
The studies followed a within-subjects design: all participants (N=6) experienced all conditions. 
In Pilots 1 and 3, we tested three different duration conditions and six different timing conditions separately. Pilot 2 examined five types of auditory feedback. As a result, for Pilots 1 and 3, we counterbalanced the conditions using a Latin Square design, while in Pilot 2, with six participants and five conditions, a Latin Square design was infeasible, so we randomized the feedback types for each participant.

\begin{figure}[hpbt]
	\centering
	\includegraphics[width=1\linewidth]{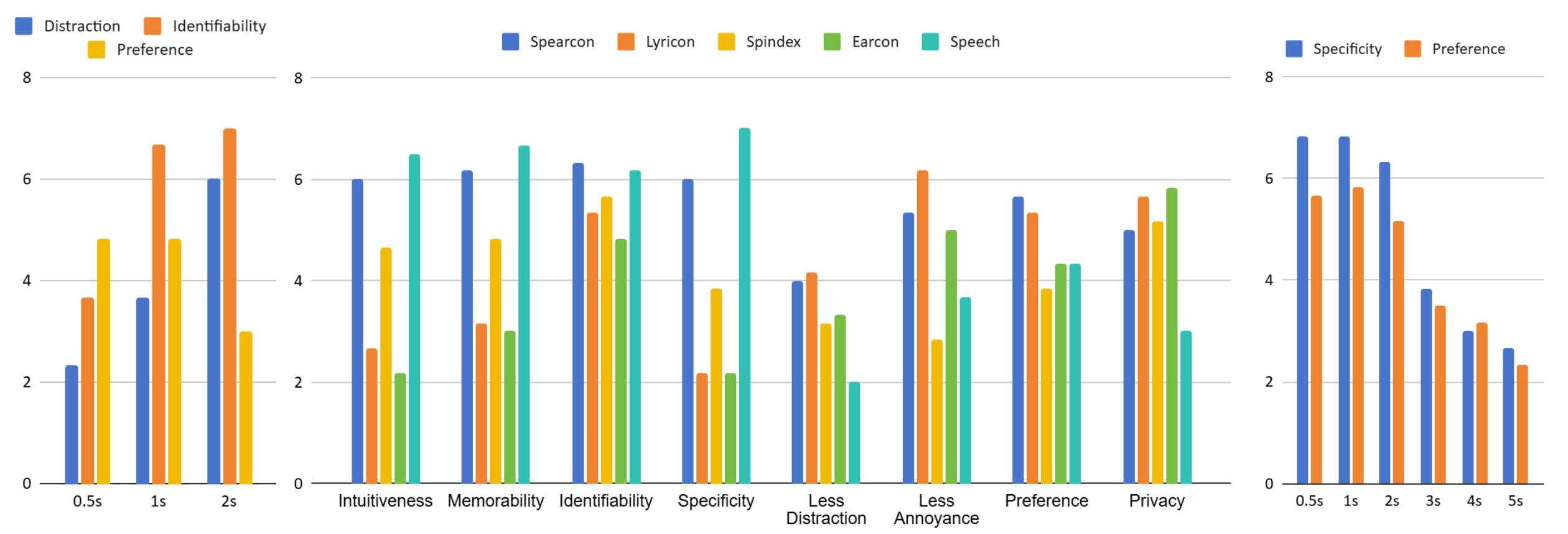}
	\caption{Participants' (N=6) ratings for the (a) Pilot 1: Duration, (b) Pilot 2: Type, and (c) Pilot 3: Timing.}
	\label{fig:pilot:evaluation_type}
\end{figure}

\subsection{Pilot 1: Feedback Duration}
\paragraph{Design}
We tested three feedback durations: 0.5s, 1s, and 2s, focusing on measuring distraction, identifiability (i.e., if the feedback is easily distinguishable from other environmental sounds, see Sec~\ref{sec:design_space}), and user preference.
Earcons were chosen for auditory feedback due to their adjustable duration \cite{blattner_earcons_1989}, in contrast to the rest, which have fixed duration (e.g, spearcon, speech). 
Different audio samples were available, and participants were able to freely map specific unwanted words to the sound they preferred.

\paragraph{Results}
As depicted in Figure~\ref{fig:pilot:evaluation_type} (a), participant feedback suggested a trend: shorter durations led to less distraction, while longer durations improved identifiability. Preferences were similar for the 0.5s and 1s durations, but the 2s duration seemed less preferred. Although the 1s duration seemed to offer a balance between minimizing distraction and ensuring identifiability, durations shorter than 1 second could still be effective for heightening awareness of unwanted words, provided that preference is maintained.

\subsection{Pilot 2: Feedback Type}
\paragraph{Design}
We evaluated five auditory feedback types: one speech-based (i.e., the utterance of the unwanted word) and four non-speech variants (earcon, spearcon, spindex, and lyricon). Due to their inability to adequately represent specific unwanted words with suitable metaphoric mappings \cite{garzonis_auditory_2009, cabral_auditory_2019}, auditory icons were not considered. The earcons were set to a duration of 1s. For Lyricon—typically ranging from 0.5 to 2 seconds—were filtered to include only variants under 1 second in duration, according to Pilot 1's results. The other feedback\footnote{Note: All non-speech feedback had durations less than 1 second, while the unwanted word determined speech feedback duration.} types adhered to their original design methodologies \cite{walker_spearcons_2013, jeon_lyricons_2013, jeon_spindex_2011}. For example, \autoref{tab:ExampleofAuditoryFeedback} shows the five auditory feedbacks for the word "like" selected by a participant. Corresponding audio samples are available as supplementary material (see \material) and in the accompanying video. For the non-speech conditions, we collected the most common non-speech auditory feedback and allowed participants to map the unwanted words they wanted to reduce with the specific auditory samples based on their selection (e.g., preferences) before the actual session.

\begin{table}[hptb]
\centering
\caption{An example of five auditory feedback types selected by a participant in Pilot 2. \significantI{} indicates feedback types that are customizable to participants and \significantII{} indicates feedback types that are common to all participants.}
\label{tab:ExampleofAuditoryFeedback}
\small
\begin{tabular}{@{}cll@{}}
\toprule
\multicolumn{1}{l}{\textbf{Unwanted Word}} & \textbf{Feedback Type} & \textbf{Example} \\ \midrule
\multirow{5}{*}{Like} & Earcon\significantI{} & A brief “buzz” sound intended to alert the user. \\
 & Lyricon\significantI{} & A musical lyric that plays "DaDi" to provide rhythmic auditory feedback. \\
 & Spindex\significantII{} & A distinct sound of the letter "L" designed to be recognizable to the user. \\
 & Spearcon\significantII{} & An auditory feedback that involves speeding up the pronunciation of the word "like.” \\
 & Speech\significantII{} & A playback of the word "like" itself, used as a direct auditory feedback. \\ \bottomrule
\end{tabular}
\end{table}

\paragraph{Results}
As shown in Figure~\ref{fig:pilot:evaluation_type} (b), participant feedback identified spearcon as the most favorable overall\footnote{An average score was calculated for all metrics.}. Speech feedback ranked second in positivity but was sometimes perceived as mocking by users due to its direct delivery, leading to occasional annoyance. Although lyricon, spindex, and earcon were perceived as less annoying and less distracting, they were perceived as less intuitive, requiring users to remember specific mappings to unwanted words.

\subsection{Pilot 3: Feedback Timing}

\paragraph{Design}
Based on prior findings, spearcon was selected as the feedback type. We tested six delay times: 0.5s, 1s, 2s, 3s, 4s, and 5s, focusing on specificity and preference as the primary measures. This is because spearcon supports higher intuitiveness, memorability, and privacy, among other factors (see previous pilot). The delays were implemented by adding a lag in the system whose duration was the tested delay minus 0.5 seconds, which corresponds to the minimum delay time in the Wizard of Oz setup, according to our tests.

\paragraph{Results}
As shown in Figure~\ref{fig:pilot:evaluation_type}(c), participant responses indicated a general trend where longer delay times led to diminished specificity and preference. Specifically, delay times above 2s significantly impacted specificity, leading to confusion among participants about the feedback's relevance to specific unwanted words, which in turn negatively affected their preference. Consequently, the results suggest that a delay time of less than 2 seconds is preferable.

\section{\WSCOACH{}: Wearable Real-Time Auditory Feedback System Design}
\label{sec:system:wsc}

Informed by insights from our pilot studies, we developed \WSCoach{} (\textbf{W}earable \textbf{S}peech \textbf{Coach}), a proof-of-concept wearable system designed to minimize the use of unwanted words in daily conversations through (near) real-time auditory feedback. While the implementation is relatively simple, \WSCoach{} provides a means to verify whether such (near) real-time feedback is feasible and effective. \WSCoach{} utilizes spearcons for feedback, chosen for their effectiveness in making unwanted words easily identifiable.

\WSCoach{} supports Bluetooth-enabled auditory devices (e.g., smart glasses) equipped with built-in microphones. We selected Huawei Eyewear (Version 1\footnote{\url{https://web.archive.org/web/20221007024433/https://consumer.huawei.com/en/wearables/huawei-eyewear/}}) due to its technical compatibility and availability during the study.
In a pilot evaluation (N=4 participants) focused on power consumption, we tested \WSCoach{}'s battery performance during full-day usage in natural conversations. The device, fully charged at the beginning, operated continuously in the background to detect filler words. The average battery life was 341.5 minutes (SD = 54.8 min, approximately 5.7 hours). The device uses Bluetooth 5.0, which supports a maximum data rate of 2 Mbps\footnote{\url{https://www.iotforall.com/bluetooth-5-iot}}, and connects to a computer using the default sub-band codec (SBC) commonly employed in wireless headphones. SBC latency typically ranges from 150 to 250 ms\footnote{\url{https://www.rtings.com/headphones/tests/connectivity/bluetooth-connection}}, depending on transmission distance and environmental conditions.

\WSCoach{} is implemented in Python (v3.9) and comprises three main modules: \textit{Record}, \textit{ASR} (automatic speech recognition), and \textit{TTS} (text-to-speech). 
The \textit{Record} module captures audio using PyAudio\footnote{\url{https://pypi.org/project/PyAudio/}} (v0.2) and interfaces with auditory device via Bluetooth. It integrates with background noise reduction tools such as NVIDIA RTX Voice\footnote{\url{https://www.nvidia.com/en-us/geforce/guides/nvidia-rtx-voice-setup-guide/}}, which employs GPU-based AI filtering to suppress background sounds during input capture. This significantly improves system robustness in noisy environments such as offices and cafés.
The \textit{ASR} module transcribes audio using the Faster-Whisper\footnote{\url{https://github.com/systran/faster-whisper}} (v0.7) multilingual speech recognition engine. While the current configuration targets detection of unwanted words in English, the GUI allows selection of other supported languages.
The \textit{TTS} module closes the feedback loop by playing pre-generated spearcons through the smart glasses upon detection of unwanted words. Spearcons are synthesized using pyttsx4\footnote{\url{https://pypi.org/project/pyttsx4/}} (v3.0) and compressed to 40\% of their original duration, following Walker et al. \cite{walker_spearcons_2013}, to ensure brevity and urgency.

The system runs on a desktop PC equipped with a GeForce RTX 3080 Ti GPU to enable low-latency processing. Source code and implementation details are available at \material.

Overall, the system achieved an average word detection latency of 0.81 seconds (SD = 0.24, range = 0.38–1.95), followed by an average spearcon playback time of 0.28 seconds (SD = 0.08, range = 0.14–0.46), resulting in a total feedback delivery time of approximately 1.1 seconds on average.

\section{Evaluation Study: Empirical Comparison between \WSCOACH{} and \ORAI{}}
\label{sec:eval}

To evaluate the effectiveness of \WSCoach{}, we ran a comparative study using the \Orai{} mobile application (\url{https://orai.com/}) as a baseline. This commercial application offers retrospective feedback on unwanted words after a conversation. As we will explain later, it represents the status-quo approach for reducing the occurrence of unwanted words. 
As we are interested in the effectiveness of each system in both short-term and longer-term usage, we included two phases for our study: a training phase and a post-training phase. These two phases are detailed later in Sec~\ref{sec:study2:design}.

\label{sec:research-questions}

\vspace{3mm}\noindent{}
The primary research questions (RQs) guiding our investigation were as follows:

\begin{itemize}[leftmargin=*]

    \item \RQI{, demonstrating short-term effectiveness}

    \item \RQII{, indicating lasting effectiveness}

    \item \RQIII{}

    \item \RQIV{}
\end{itemize}

Furthermore, we were interested in exploring the following secondary (exploratory) research questions:

\begin{itemize}[leftmargin=*]

    \item \RQV{}

    \item \RQVI{}
\end{itemize}

We pre-registered our experiment design, research questions, hypotheses, and analysis plan in order to increase the reliability of our findings \cite{nosek_preregistration_2018} (see \material).

\subsection{Participants}

We recruited 24 participants (P1-P24, 12 male, 12 female, mean age = 23, SD = 3.5) from our university, none of whom took part in our pilot studies. All participants reported in a recruitment questionnaire that they have no auditory or visual impairment, that they want to reduce unwanted words in their daily conversations, and that they have professional-working fluency in English.
They were compensated 7.31 USD per hour for their time.

The sample size of $N$=24 was chosen ahead of time, considering that it would give us a power of 0.7 to detect a standardized effect size of $d$=1 (conventionally referred to as a large effect size) using an independent-sample t-test, as calculated with G*Power. Although such a sample size is insufficient to detect small differences in effectiveness between our two experimental conditions, we were constrained by the time and financial costs of running participants.

\subsection{Apparatus}
\label{sec:study2:apparatus}

The study required participants to converse with the experimenter using either the \WSCoach{} (Sec~\ref{sec:system:wsc}) or the \Orai{} system. For the \Orai{} system, conversations were facilitated using the \Orai{} mobile app on a Redmi K40 smartphone. All conversations were audio-recorded with a Huawei P10 to allow for post-analysis.

\Orai{} was selected as the baseline of comparison due to its reputation as an effective AI speech coach, offering comprehensive training in speaking skills. 
It is a highly popular app, evidenced by over 100,000 downloads and a 4.1-star rating on Google Play. The app features post-conversation analysis, including audio transcriptions highlighting unwanted words, their frequency, and playback capability for in-depth self-evaluation (See Appendix~\ref{appendix:orai_app}).

\subsection{Study Design}
\label{sec:study2:design}

\begin{figure*}[tbh]
	\centering
	\includegraphics[width=1\linewidth]{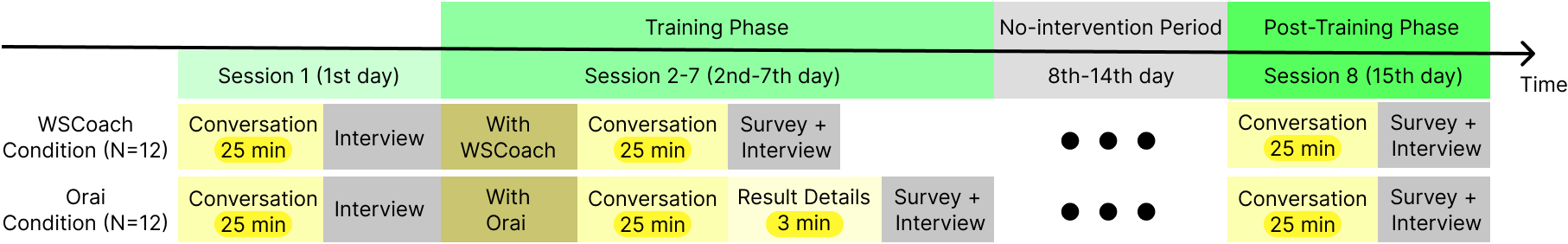}
	\caption{Experiment design used in the comparative evaluation, showing the sequence of experimental sessions for the \WSCoach{} experimental group (top row) and the \Orai{} group (bottom row).}
	\label{fig:study2:procedure}
\end{figure*}

The study followed a between-subjects design. Participants were randomly assigned to one of two groups: \WSCoach{} or \Orai{}. This assignment was based on a pre-generated, shuffled list that determined the condition of each participant according to the order in which they completed the recruitment questionnaire.

All conversation sessions were conducted in a controlled laboratory setting, ensuring minimal environmental interference such as noise or other distractions. All participants adhered to their scheduled times and completed the entire experiment.

Each participant engaged in eight conversation sessions across 15 days, for a total of about four hours of participation. The sequence of sessions is summarized in Figure~\ref{fig:study2:procedure} for each experimental condition. We provide a brief overview here, with more details in the following subsection. The first conversation took place without any technology assistance. Then, the training phase consisted of six conversation sessions with the system (\WSCoach{} or \Orai{}), on consecutive days.
This number was chosen based on prior studies that apply behavioral interventions across six sessions \cite{havighurst2010tuning, fischer2011cross}. Each conversation lasted about 25 minutes, measured by a timer starting at the beginning and stopping once the time had elapsed, recording the total duration. This choice was inspired by a well-known time management method that recommends working in 25-minute segments \cite{cirillo2018pomodoro}. 
The training phase was followed by a no-intervention period during which participants did not use any system. Finally, participants attended a final session to assess their improvement.

\subsection{Task and Procedure}
We now go through the phases shown in Figure~\ref{fig:study2:procedure} and detail the procedure.

\textbf{First session}. After giving consent and receiving the briefing, the participant was seated at a table (see Figure~\ref{fig:study2:setting}) and had a first conversation with the experimenter (one co-author, always the same), without any technological assistance. The goal of this first session was to establish a baseline count of unwanted words, and to help participants familiarize themselves with the settings.
The conversation was guided by the experimenter, based on topics derived from IELTS Speaking Questions (\url{https://ielts.org/}), which cover a range of familiar subjects. Each session used a different set of topics (see \material). The experimenter assumed a facilitator role by listening attentively, aligning with participants' viewpoints, and steering the conversation when it began to stall. They were also trained to avoid speaking unwanted words and to suspend system detection when the participant was not speaking.
After the conversation, the participant was interviewed to identify which unwanted words they wished to reduce. 

\textbf{Training phase}. This phase took place from day 2 to day 7. 
On the second day, the participant was introduced to the system that was randomly assigned to them (\WSCoach{} or \Orai{}). After familiarization with the system, the participants engaged in a conversation session, assisted by the system. They kept using the same system for the next five conversation sessions (days 3 to 7).  
In the \WSCoach{} group, participants put on smart audio glasses with their lenses removed before each conversation. Those already wearing spectacles were invited to remove them or place smart audio glasses over them. 
In the \Orai{} group, participants were handed a smartphone with \Orai{} installed, and were invited to open the app and place the phone on the table to monitor the conversation. They then had to spend at least 3 minutes reviewing their use of unwanted words with the \Orai{} application. Additionally, in both groups, participants completed a questionnaire and participated in an interview about their experiences and the system's pros and cons (see Sec~\ref{sec:study2:measures} on Measures for more details). In the last training session (the 7th session), participants completed an additional questionnaire and participated in a 15-minute interview about their perceived improvement during the training phase. The complete training phase interview questions are available in \material{}. 

\textbf{No-intervention period}. The training phase was followed by an 8-day break during which participants were not involved in any sessions and did not use any system.

\textbf{Post-training phase} On the 15th day, participants engaged in their final conversation with the experimenter. Afterward, they completed a post-training phase questionnaire and were interviewed about their perceived improvement. The complete post-training phase interview questions are available in \material{}

\subsection{Post-Experiment Data Processing and Analysis}
\label{sec:blind-coding}

\paragraph{Quantitative Coding} Post-experiment, all conversation recordings and post-interview recordings were anonymized and transcribed with the help of Deepgram (\url{https://deepgram.com/}). 
We reviewed all transcripts for accuracy and redacted any content that might disclose the experimental condition (\WSCoach{} or \Orai{}), for the purpose of the subsequent blind coding.
Two independent coders, who did not attend the interviews and were naive about the purpose of the research, were provided with the anonymized conversation transcripts and a list of unwanted words for each conversation (designated by the participant in the first session). The coders were tasked with counting the occurrences of unwanted words in each conversation and measuring speaking time, i.e., the amount of time the participant speaks.
We assessed inter-coder agreement for the counts of unwanted words and the amount of time participant speaks using Cohen's Kappa \cite{cohen1960coefficient}, obtaining $\kappa =0.85$ and $\kappa =0.81$ separately, both reflecting high agreements. Disagreements were resolved through discussion between the two coders.

\paragraph{Qualitative Coding} 
Moreover, we used thematic analysis \cite{braun2006using} to analyze the qualitative data (e.g., interview). First, we (two co-authors) reviewed the qualitative data records multiple times. Then, we identified the participants' responses, grouped the answers, and incorporated them into themes and sub-themes. We finalized our themes and discussed the different speech intervention systems in detail, along with guidelines for real-time auditory intervention. We report the qualitative results later in the paper.

\subsection{Measurements}
\label{sec:study2:measures}

\begin{table*}[t]
\centering
\caption{The measurements used in Study 2.}
\label{tab:Study2Measurements}
\renewcommand\arraystretch{1.2}{ 
\small
\setlength{\extrarowheight}{-2pt} 
\begin{NiceTabular}{llll}
    \toprule
    \multirow{2}{*}{\vspace{3mm}\textbf{Independent}} & \multirow{2}{*}{\textbf{Platform}} & \multirow{1}{*}{$\bullet$ Glasses} & \multirow{1}{*}{$\bullet$ Smartphone}\\
    \textbf{Variables} & & Using \WSCoach{} for reducing unwanted words & Using \Orai{} for reducing unwanted words\\
    \cmidrule(lr){1-4}
    \multirow{2}{*}{\vspace{4mm}\textbf{Measurements}} & \multirow{2}{*}{\vspace{3mm}\textbf{Primary }} & \multicolumn{2}{l}{\textit{$\bullet$ Improvement during the training phase,} measured as the rate of change in the normalized}\\
    \textbf{\vspace{3mm}(Dependent} & \textbf{Measurements} & \multicolumn{2}{l}{ frequency of unwanted words $S_n$, from the 1st to the 7th session (see Sec~\ref{sec:study2:measures}).}\\
    \textbf{Variables)} &  & \multicolumn{2}{l}{$\bullet$ \textit{Improvement at the post-training phase}, defined as the normalized frequency of unwanted}\\
     &  & \multicolumn{2}{l}{words at session 8 (Sec~\ref{sec:study2:measures}).}\\

    \cmidrule(lr){2-4}
     & \multirow{2}{*}{\vspace{4mm}\textbf{Secondary}} & \multicolumn{2}{l}{\textit{$\bullet$ Training phase -- Confidence in Speaking} (1-7): "This system made me feel more confident}\\
    & \textbf{Measurements} & \multicolumn{2}{l}{in my conversations."}\\
    & & \multicolumn{2}{l}{\textit{$\bullet$ Training phase -- Awareness of Speaking Unwanted Words} (1-7): "This system helped me }\\
     & & \multicolumn{2}{l}{ become more aware of my unwanted words during conversations."}\\
    &  & \multicolumn{2}{l}{\textit{$\bullet$ Training phase -- Self-rated Conversation Quality} (1-7): "My quality of conversation has}\\
    &  & \multicolumn{2}{l}{improved with the help of this system."}\\

    &  & \multicolumn{2}{l}{\textit{$\bullet$ Training phase -- 
 Distraction of Feedback} (1-7): "The system distracted me from having a}\\
    &  & \multicolumn{2}{l}{natural conversation."}\\

    &  & \multicolumn{2}{l}{\textit{$\bullet$ Training phase -- Raw NASA TLX} (0-100) for Being Aware of Unwanted Words}\\

         &  & 
     \multicolumn{2}{l}{\textit{$\bullet$ Post-training phase -- Confidence in Conversation} (1-7): "I felt confident in maintaining the}\\
     & & \multicolumn{2}{l}{reduction of unwanted words without the aid of the system."}\\
     &  & \multicolumn{2}{l}{\textit{$\bullet$ Post-training phase --Awareness of Avoiding Unwanted Words} (1-7) = "I found myself }\\
    &  & \multicolumn{2}{l}{consciously avoiding unwanted words."}\\

     \cmidrule(lr){1-4}

    {\vspace{4mm}\textbf{Supplementary}} &  & \multicolumn{2}{l}{\textit{$\bullet$ Normalized Ratio of Unwanted Words}: The total number of unwanted words divided }\\
    \textbf{Measurements}  &  & \multicolumn{2}{l}{ by the total number of words participant spoke.}\\

    \bottomrule
\end{NiceTabular}}
\end{table*}

The effectiveness of reducing unwanted words and the quality of conversation were evaluated using objective and subjective measurements, as summarized in Table ~\ref{tab:Study2Measurements}. In addition, participants' perceptions of real-time auditory feedback and delayed memory were captured in the interviews.

To help address statistical multiplicity \cite{bender2001adjusting, li2017introduction}, we separate our measurements into \textit{primary measurements}, which will serve to answer our main research questions, and \textit{secondary measurements}, which are more exploratory. Additionally, we introduced a set of \textit{supplementary measurements}, which were not pre-registered and added after peer review, to address reviewers' questions and comments.

\subsubsection{Primary Measurements.} We have two primary measurements, both of which are defined based on a metric we call the \textit{normalized frequency of unwanted words}. We refer to the normalized frequency of unwanted words in session $n$ as $S_n = \dfrac{F_n}{F_1}$, with $1 \leq n \leq 8$, and $F_n$ being the \textit{frequency of unwanted words} in session $n$, defined as the ratio between the number of unwanted words in session $n$ and the participant's total speaking time in that session. Accordingly, $S_1 = 1$ for all participants. Our first primary measurement is the rate of change in $Sn$ from the 1st to the 7th session. The way the seven measurements of normalized frequency (one per session) are aggregated into a single rate measurement will be explained in more detail in the analysis section. This measurement captures the participant's improvement during the training phase and will serve to answer research questions RQ1 and RQ3 (see Sec~\ref{sec:research-questions}). Our second primary measurement is $F_8$, which captures the long-term improvement as measured in the post-training phase, and it will serve to answer RQ2 and RQ4.

\subsubsection{Secondary Measurements.} For the training phase, we evaluated several aspects to determine \WSCoach{}'s impact on conversation quality and user experience. We recorded \textit{Self-rated Conversation Quality} to see if the system made conversations feel unnatural. We measured the \textit{Distraction of Feedback} to assess if participants were distracted by the system. Frequent auditory feedback might negatively affect participants' confidence, so we measured \textit{Confidence in Speaking}.” Awareness of unwanted words indicates participants' improvement, so we measured \textit{Awareness of Speaking Unwanted Words}. All these metrics were measured on a 7-point Likert scale. Additionally, we evaluated the cognitive load using the perceived task load (NASA TLX) \cite{hart2006nasa} to assess the workload involved. For the post-training phase, we collected  \textit{Confidence in Conversation} to evaluate the system's lasting effect on users' confidence and \textit{Awareness of Avoiding Unwanted Words} to measure the lasting self-awareness regarding unwanted words. These measurements in the training and post-training phases will serve to answer RQ5 and RQ6.

\subsubsection{Supplementary Measurements.}
We calculated the \textit{Normalized Ratio of Unwanted Words} as an alternative metric to analyze the improvement during the training phase and the post-training phase. For each session $n$, it is defined as $S_n' = \dfrac{R_n}{R_1}$, with $1 \leq n \leq 8$, and $R_n$ being the ratio of unwanted words in session $n$, defined as the ratio between the number of unwanted words in session $n$ and the number of participant's total words in that session. This new metric has the advantage of not being affected by changes in speech rate.

\subsection{Statistical Analysis}

\subsubsection{Planned Analysis.}
The analyses reported here were planned and preregistered before data collection (see \material).

\paragraph{\RQI{}}
For each participant in the \WSCoach{} group, we performed a linear regression with normalized frequency $S_n$ as the dependent variable and session number $n \in [1..7]$ as the independent variable. We used the regression slope to measure the rate of change in unwanted words across the training phase for each participant. We then computed the mean slope for the entire \WSCoach{} group and performed a one-sample $t$-test with zero (horizontal slope, meaning no change overall) as the null hypothesis. From this $t$-test we derived a $t$-based 95\% confidence interval for the mean slope and a $p$-value, capturing the amount of evidence in favor of an overall reduction of unwanted words during the training phase.

\paragraph{\RQII{}} 
We performed a one-sample $t$-test on the mean normalized frequency $S_8$ (normalized frequency at the 8th session, i.e., after about 15 days) for the \WSCoach{} group, with 1 as the null hypothesis (meaning no change). This $t$-test allowed us to assess whether there was a reliable overall improvement after the no-intervention period compared to the very first session.

\paragraph{\RQIII{}} For each participant in the \Orai{} group, we performed the same linear regression as we did for the \WSCoach{} group (see RQ1). We then performed an independent-sample \textit{t}-test to see if there is a reliable difference between the mean slopes of the two groups.

\paragraph{\RQIV{}} We performed an independent-sample \textit{t}-test to compare the mean $S_8$ between the \WSCoach{} group and the \Orai{} group.

\subsubsection{Secondary Analyses of Primary Measurements.}
In this subsection and all following subsections, all analyses are post-hoc and not pre-registered. Therefore, any finding should be taken as tentative.

\paragraph{Normalized Frequency S7.} Our regression method is one way of operationalizing improvement in the learning phase (RQ1), but another way is to compare the last training session with the initial session, i.e., by looking at the normalized frequency $S_7$. We performed an independent-sample \textit{t}-test to compare the mean $S_7$ between the two groups.

\paragraph{Robustness Check Using S'} In addition, we introduced a new metric for measuring the occurrence of unwanted words: $S_n'$, the \textit{Normalized Ratio of Unwanted Words}, defined in Sec~\ref{sec:study2:measures}. As a robustness check, we re-analyzed our data using $S_n'$ instead of $S_n$ and re-assessed our primary research questions (RQ1 to RQ4).

\subsubsection{Analyses of Secondary Measurements.}
For all Likert items, we computed the mean response (range 1--7) and its \textit{t}-based 95\% CI for each group. We also computed the \textit{p}-value for the Mann-Whitney $U$ test of the difference between groups.

All statistical analyses were conducted using SPSS (Statistical Package for the  Social Sciences).

\section{Evaluation Study: Results}
\label{sec:evalResult}

Overall, both \WSCoach{} and \Orai{} helped reduce participant-identified unwanted words. While \WSCoach{}'s real-time feedback caused more distraction than \Orai{}'s retrospective approach, it led to greater reductions in unwanted words and higher user confidence. We report our findings below, distinguishing between pre-registered and exploratory analyses in line with methodological guidelines \cite{li2017introduction}.

\subsection{Results of the Planned Analysis.}
We report \emph{p}-values but interpret them as continuous measures of evidence, using .05 as a rough landmark instead of a cut-off \cite{besanccon2019continued}.

\begin{figure*}[hptb]
	\centering
    \includegraphics[width=0.9\linewidth]{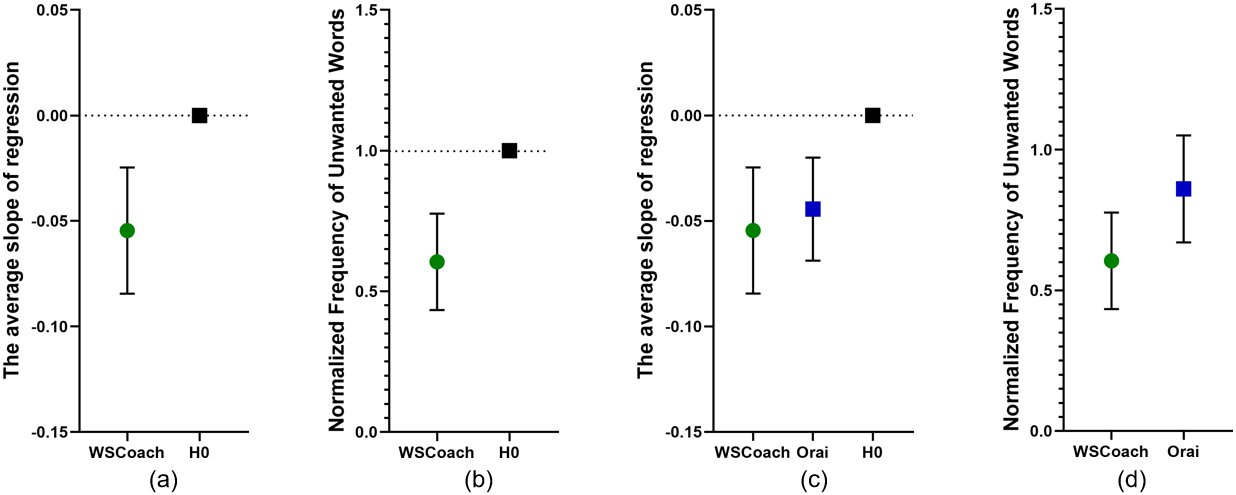}
	\caption{The planned analyses of primary measurements for the evaluation study. Lower is better. Error bars are 95\% confidence intervals. Black squares and dashed lines show the null hypothesis.}
	\label{fig:Short phase}
\end{figure*}

\paragraph{\RQI{}} 
As shown in \autoref{fig:Short phase} (a), the mean slope was $M=-0.05$\footnote{For all our reported slope values, the dependent variable is $S_n$ (unitless), and the independent variable is the session. Therefore, the slopes represent the change in the unwanted word ratio per session, but there are no units following them.}, 95\% CI [-0.08, -0.03], $t(11)=4.02$, $p=.002$. Therefore, we have strong evidence that the \WSCoach{} led to an overall improvement during the training phase.

\paragraph{\RQII{}} 
As shown in \autoref{fig:Short phase} (b), the mean normalized frequency was $M = 0.60$, 95\% CI [-0.57, -0.22], $t(11)=5.07$, $p=.0004$. Therefore, we have strong evidence that \WSCoach{} was effective and lasted up to the post-training phase.

\paragraph{\RQIII{}} 
As shown in \autoref{fig:Short phase} (c), the difference in mean slopes (\Orai{}$-$\WSCoach{}) was 0.01, 95\% CI [-0.03, 0.05], $t(22)=0.58$, $p=.57$. Therefore, we have no evidence that \WSCoach{} outperformed \Orai{} during the training phase. Additionally, the mean slope and its 95\% confidence interval for each of the two groups show that the effectiveness estimates are comparable (i.e., \Orai{} is effective as well).

\paragraph{\RQIV{}} 
As shown in \autoref{fig:Short phase} (d), the difference between the two means (\Orai{}$-$\WSCoach{}) was 0.26, 95\% CI [0.01, 0.50], $t(22)=2.2$, $p=.04$. Therefore, we have evidence that \WSCoach{} outperformed \Orai{} in the post-training phase. Additionally, the mean normalized frequencies and their 95\% confidence intervals suggest that \WSCoach{} is likely more effective in the long run (the lower the value, the better).

\subsection{Results of the Secondary Analyses of Primary Measurements}

\begin{figure*}[hptb]
	\centering
     \includegraphics[width=0.95\linewidth]{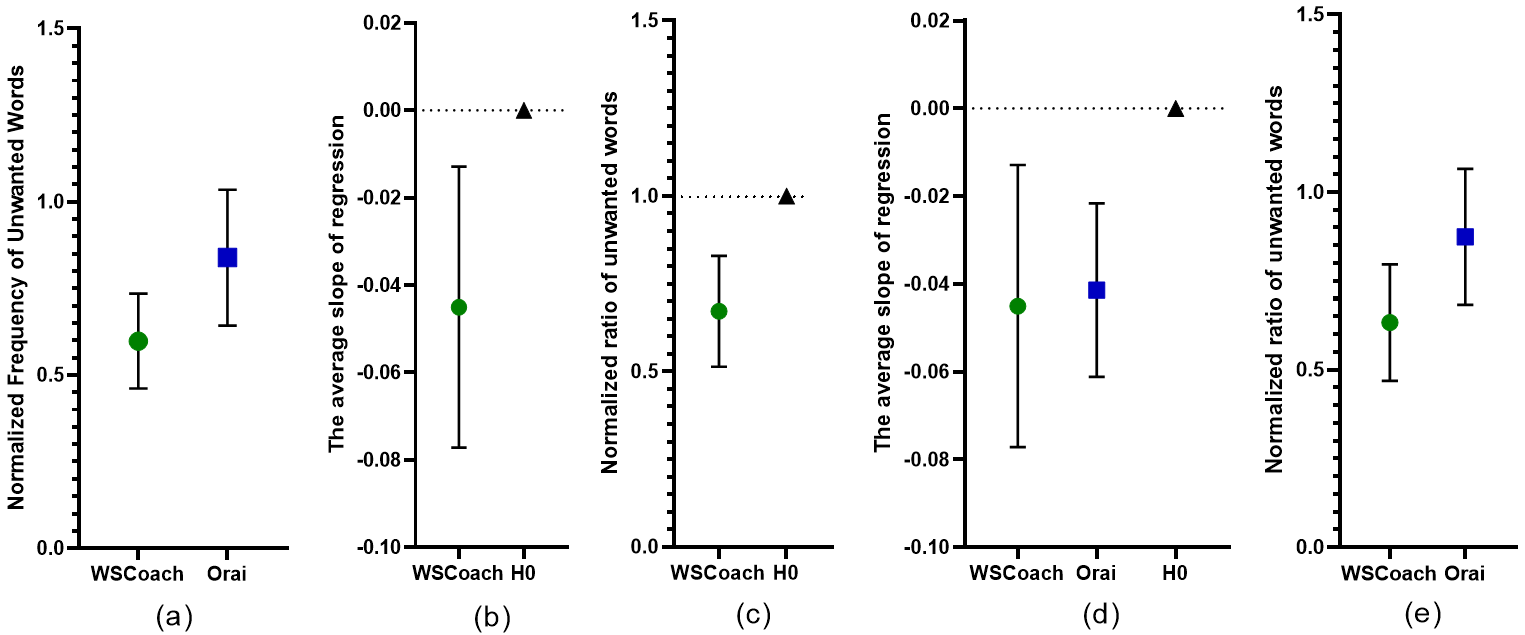}
	\caption{The secondary analyses of primary measurements for the evaluation study. Lower is better. Error bars are 95\% confidence intervals. Black triangles and dashed lines show the null hypothesis.}
	\label{fig:Additional Analysis}
\end{figure*}

\paragraph{Normalized Frequency S7.} 
As shown in \autoref{fig:Additional Analysis} (a), the difference between the two means (\Orai{}$-$\WSCoach{}) was 0.24, 95\% CI [0.02, 0.47], $t(22)=2.2$, $p=.04$. Therefore, this way of analyzing improvement during the learning phase might be more sensitive, but a replication is needed to confirm it.

\paragraph{Robustness Check Using S'} 
The conclusions are unchanged. Results are reported in \autoref{fig:Additional Analysis} (b)--(e), and can be compared to our original results in \autoref{fig:Short phase} (a)--(d):

\paragraph{RQ1:} 
As shown in \autoref{fig:Additional Analysis} (b), the mean slope was $M = -0.05$, 95\% CI [-0.08, -0.01], $t(11) = 3.09$, $p = .01$. Therefore, we still have strong evidence that the \WSCoach{} led to an overall improvement during the training phase.

\paragraph{RQ2:} 
As shown in \autoref{fig:Additional Analysis} (c), the mean normalized ratio was $M = 0.67\%$, 95\% CI [-0.49, -0.17], $t(11) = 4.57$, $p = .0008$. Thus we retain the strong evidence that \WSCoach{} was effective up to the post-training phase.

\paragraph{RQ3:} 
As shown in \autoref{fig:Additional Analysis} (d), the difference in mean slopes (\Orai{}-\WSCoach{}) was 0.004, 95\% CI [-0.03, 0.04], $t(11) = 0.21$, $p = .83$. Therefore, like before, we have no evidence that \WSCoach{} outperformed \Orai{} during the training phase. Additionally, the mean slope and its 95\% CI for each of the two groups show that the effectiveness estimates are comparable (i.e., both methods are effective).

\paragraph{RQ4:} 
As shown in \autoref{fig:Additional Analysis} (e), the difference between the two means (\Orai{}-\WSCoach{}) was 0.24, 95\% CI [0.003, 0.48], $t(22) = 2.1$, $p = .05$. Therefore, we still have evidence that \WSCoach{} outperformed \Orai{} in the post-training phase.

\subsection{Results of the Secondary Measurements.}
\label{sec:secondary-measurements}

\begin{figure} [hptb]
	\centering
 \includegraphics[width=1\linewidth]{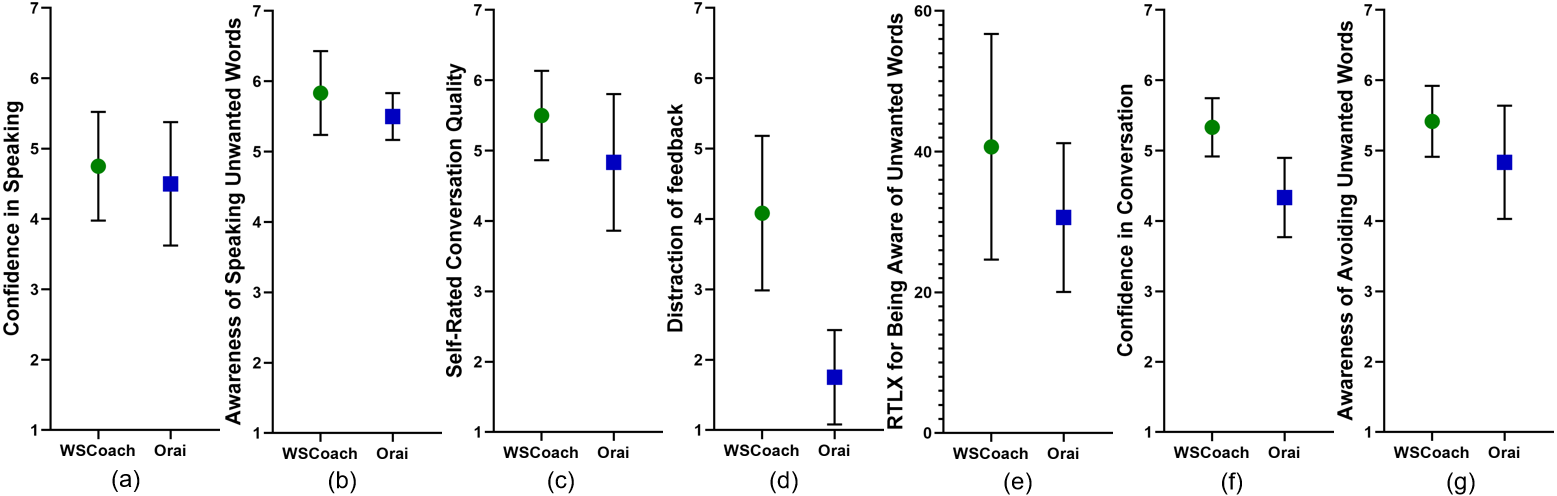}
	\caption{The analyses of secondary measurements for the evaluations study. Higher is better, except for (d) and (e), where lower is better. Error bars show 95\% confidence intervals.}
	\label{fig:Long phase}
\end{figure}

Results are reported in \autoref{fig:Long phase} (a) to (g). Here, we report the full results for the two questions for which we have evidence of a difference---the full analysis is available in \material.

\paragraph{Distraction of Feedback.} 
As shown in \autoref{fig:Long phase} (d), the mean score was 4.1, 95\% CI  [3.0, 5.2] for \WSCoach{} and 1.8, 95\% CI [1.1, 2.4] for \Orai{}, $U = 117$, $p=.003$. Therefore, we have good evidence that \WSCoach{} generated more distraction on average than \Orai{} during conversations.

\paragraph{Confidence in Conversation (post-training phase).} 
As shown in \autoref{fig:Long phase} (f), the mean score was 5.3, 95\% CI  [4.9, 5.6] for \WSCoach{} and 4.3, 95\% CI [3.8, 4.9] for \Orai{}, $U = 28$, $p=.01$. Therefore, we have evidence that \WSCoach{} made people feel more confident in their conversation than \Orai{} after the training was over.

\subsection{User Feedback and Discussion}

Based on the interview recording analysis, we discuss the user experience on two systems.

\paragraph{Established a Self-Monitoring Mechanism.} 
Participants' feedback suggests that \WSCoach{} helped establish an active self-monitoring mechanism during the training phase which helped users in reducing unwanted words: \quoteby{P7}{Whenever I hear the auditory feedback, I consciously allow myself to slow down and reflect on my words. I integrate the auditory feedback and the instruction that advises me to slow down unconsciously when approaching the expression of unwanted words during conversations.} Furthermore, our study shows that 25-minute training sessions over a 6-day period were sufficient to have a 40\% reduction in unwanted words on average in the post-training phase.

\paragraph{Improved Awareness and Lasting Impact.} 
Interestingly, after 3-4 sessions of using \WSCoach{}, participants noted improved awareness during their daily communication even when not using \WSCoach{}, highlighting the potential of \WSCoach{} to yield lasting improvements. For instance, Participant P13 mentioned on their 4th day that, \quote{During daily conversations, I have started noticing instances where I use unwanted words, prompting me to pay extra attention to monitor these words…when I'm about to utter these words, I feel a distinct `real-time auditory feedback' echoing those words in my mind… it [\WSCoach{}] has helped develop a pattern to avoid speaking those unwanted words.} Such behavioral changes are a result of operant conditioning \cite{staddon2003operant}, which posits that behaviors followed by negative consequences are less likely to be repeated. In our case, the real-time auditory interventions of \WSCoach{} acted as negative feedback – an unfavorable consequence contingent on the use of unwanted words. When users speak more unwanted words, they receive increased auditory feedback, creating a negative reinforcement loop. P2's preference for reduced auditory feedback underscores a common inclination to minimize negative consequences by reducing the utterance of unwanted words, \quoteby{P2}{When I initiate a conversation, I consciously remind myself to aim for fewer auditory feedback instances. I prefer the system to detect fewer unwanted words in my speech. Experiencing more auditory feedback makes me feel uneasy, as it suggests a lack of improvement in my communication habits.}

As highlighted earlier, our results indicate that \WSCoach{} outperforms retrospective feedback (i.e., \Orai{}) after the training is over. The higher benefits of \WSCoach{} can be primarily attributed to \WSCoach{} pinpointing instances of unwanted words in real-time and enhancing the awareness of unwanted words during conversations. P8 expressed, \quote{\WSCoach{} tells me what unwanted words I spoke, which helps me identify the locations of unwanted words in my speech during conversations. This enables me to proactively minimize them, structuring my sentences more deliberately before speaking.} In contrast, the retrospective feedback offered by \Orai{} post-conversation was found to be less actionable, resulting in a diminished immediate impact and a weaker context connection for users regarding the identified unwanted words. In addition to the real-time awareness, \WSCoach{} affords opportunities to users to proactively practice reducing the occurrence of unwanted words, in contrast to retrospective feedback, \quoteby{P1}{I am aware of the number of unwanted words I used after the conversation [using \Orai{}], but I believe it's not enough. It doesn't guide me on how to reduce the unwanted words while conversing.}

\paragraph{Impact on Distraction and Cognitive Load.} 
One important aspect of \WSCoach{} is its variable impact on perceived distraction during conversations, particularly in the early stages of use. While most participants (8/12) reported acclimating to the auditory feedback by the third or fourth day—eventually no longer finding it disruptive—this was not universal. Notably, one participant (P19) consistently found the system distracting, which paradoxically led to an increase in their use of unwanted words. As they described: \quoteby{P19}{I believe the reason this intervention might not be friendly to me is that my thoughts tend to flow continuously, and I can be quite sensitive to negative feedback. So, when I'm interrupted, I need to reorganize my thoughts; in such situations, my use of unwanted words might increase.} This experience suggests that while real-time feedback can raise awareness and encourage behavior change, it may also disrupt cognitive flow—particularly for users who are sensitive to interruption or who rely on continuous verbal expression. For such individuals, real-time feedback may inadvertently increase cognitive load and hinder performance, at least initially.

Cognitive load is a critical factor in behavior change interventions, especially when targeting deeply ingrained habits \cite{kwasnicka2016theoretical}. \WSCoach{} introduces some initial effort: users must attend to both their speech and the system’s feedback, which can momentarily disrupt the natural rhythm of conversation. Nonetheless, most participants found that this disruption subsided after 2–3 days, as the feedback became less intrusive and more intuitive. This adaptation curve aligns with habit formation research, which emphasizes the role of repetition and contextual cues in shifting behaviors from effortful to automatic. As Lally et al. \cite{lally2010habits} note, such transitions often unfold over weeks or months, highlighting the need for sustained support. In our study, both \WSCoach{} and the control condition (\Orai{}) led to short-term reductions in unwanted word use. However, these gains tended to diminish by Day 15 when the intervention was withdrawn, suggesting that while immediate benefits are achievable, long-term change may require ongoing or periodic reinforcement.

Ultimately, we view the initial cognitive load not as a deterrent, but as a natural part of the behavior change process. \WSCoach{} is designed to minimize disruption through brief, subtle spearcons, and our findings suggest it is generally well tolerated. Still, long-term usability and cognitive effort remain important considerations.

This diversity in user responses highlights the need to account for individual differences in the design and application of real-time auditory interventions. A viable solution could involve adaptive features allowing users to tailor the feedback's intensity or frequency to their preferences, thus harmonizing the intervention with their cognitive processes. Additionally, a gradual feedback intensification strategy over initial sessions may facilitate a smoother adaptation, preventing users from feeling overwhelmed and promoting gradual adjustment. Personalizing interventions is key to accommodating diverse user needs, thereby maximizing efficacy and minimizing potential distractions.

\section{General Discussion}

The \WSCoach{} prototype demonstrates how immediate auditory intervention during conversational moments can effectively modify habitual speech behaviors in everyday interactions. Building such real-time intervention systems requires careful consideration of feedback types and real-time auditory intervention design. Below, we outline key aspects that future systems should consider and refine for effective digital speech behavior interventions.

\subsection{Real-Time vs. Retrospective Feedback for Speech Intervention}

Our research highlights the impact of real-time auditory feedback on effective speech-behavioral intervention. By delivering immediate auditory feedback, users become more aware of specific speech patterns, enabling them to recognize and address unwanted behaviors in context. This immediacy allows for quick reflection and self-correction, creating a reinforcement loop that promotes sustainable speech behavior modification. During our experiments, participants responded to auditory feedback by considering recently used unwanted expressions and adjusting their speech accordingly. This approach contrasts with retrospective feedback, such as that offered by public speaking tools like \Orai{}, where users can review and analyze their performance post-speech in a focused environment. While useful for skill-building in structured, single-focus tasks, retrospective feedback lacks the immediacy essential for spontaneous speech correction within interactive dialogue. In dynamic, multitasking environments, real-time feedback is uniquely suited for enabling users to identify and regulate behaviors such as filler word overuse without interrupting conversational flow. This approach aligns well with the demands of natural conversations, enhancing real-time communication skills by encouraging self-awareness and facilitating rapid improvements.

\subsection{Privacy Considerations for Real-Time Speech Intervention Systems}

Privacy is a central concern in the design and deployment of real-time speech feedback systems like \WSCoach{}. Although the system processes only \textbf{audio-only} input (i.e., no visual data), its always-on nature and operation in social environments raise important privacy considerations \cite{datta2018survey} for both users and bystanders.

\paragraph{User Privacy}
\WSCoach{} is designed to perform speech detection in real time and discard raw audio immediately after processing. Only minimal metadata (e.g., frequency of unwanted word usage) is retained. Nevertheless, several participants (3/12) expressed concerns about the possibility of sensitive speech being unintentionally processed, even when the \textbf{audio itself is not stored}. While \WSCoach{} avoids long-term recording, its continuous operation still led some users to perceive it as a form of surveillance. However, real-time auditory feedback was generally regarded more favorably than retrospective, phone-based recordings. More participants (6/12) viewed the latter as more invasive due to the potential for long-term storage and external access.

As edge computing capabilities continue to advance, future iterations may shift more of the processing directly onto the glasses themselves, further reducing the need for data transmission and minimizing the risk of data exposure \cite{zhou_edge_2019, mendez_edge_2022}. Additionally, user-facing privacy controls—such as those modeled on Apple’s App Tracking Transparency framework \cite{inc_user_2024}—can enable fine-grained control over when processing occurs and what data is retained.

Moreover, several participants (4/12) expressed a preference for subtle auditory feedback to avoid drawing attention from bystanders and to maintain social appropriateness. These insights underscore the importance of delivering feedback discreetly and in a privacy-preserving manner—ideally through speaker-only audio using bone conduction or directional sound.

\paragraph{Bystander Privacy}
While users may consent to speech processing, bystanders may not. This introduces a distinct ethical challenge. One participant (P14) specifically noted discomfort using the system around friends or family, expressing concern that it could inadvertently process sensitive portions of others’ conversations. Unlike some mobile apps that store entire conversations, \WSCoach{} neither stores audio nor attempts to associate it with specific individuals. Nevertheless, the mere presence of a listening device can raise concerns about passive surveillance.

To address this, future versions of \WSCoach{} will explore integrating speaker diarization and voice activity detection techniques to isolate the wearer’s voice. Combined with spatial filtering through multi-microphone arrays in smart glasses, these techniques—based on recent advances in wearable voice segmentation and beamforming—could help ensure that feedback is triggered only by the user’s speech. This capability is crucial for maintaining bystander trust and enabling ethical, socially acceptable deployment.

By limiting data collection, enabling on-device processing, and supporting selective voice detection, \WSCoach{} aims to balance utility with strong privacy safeguards. These considerations are essential not only for user adoption but also for the responsible integration of speech feedback technologies into everyday social contexts.

\subsection{What Kind of Real-Time Feedback Should be Provided?}
Participants using \WSCoach{} highlighted its effectiveness in helping them recall and reduce the use of unwanted words during conversations. Prior studies have shown that detailed feedback enhances knowledge retention and the application of knowledge \cite{wojcikowski2013immediate}. The design of real-time auditory feedback is pivotal in achieving this, demanding a careful balance between the richness of information, duration, and minimization of distraction during conversations. Additionally, balancing frequent practice with user endurance is essential when designing real-time auditory feedback systems that foster lasting behavior change.

\subsubsection{Trade-off: Detailed Information and Duration} 
The evaluation study (Sec~\ref{sec:evalResult}) has shown that auditory feedback with sufficient information effectively reduces various unwanted words during live conversations. However, while providing detailed information, the duration of auditory feedback increases, which may disturb users. Our pilot study results suggest that feedback should be detailed enough to convey information about specific unwanted words spoken, ensuring it remains meaningful. Nonetheless, it should also be limited to a maximum duration of 1 second to allow users to quickly comprehend it without disrupting the flow of conversation. This balance can be achieved through the careful selection and prioritization of spearcon content for different unwanted words. Managing this trade-off is essential for optimizing the auditory feedback system's effectiveness and user experience.

\subsubsection{Trade-off: Detailed Information and Distraction}
We observed a slight distraction while using \WSCoach{} due to the reception of detailed information with real-time auditory feedback during fluent conversations. Our pilot studies found that other non-speech-based feedback is less distracting than spearcons; however, spearcons provide more informative feedback. Notably, the majority of participants (11/12) valued the detailed feedback. Although it initially caused some distraction, most adapted within three sessions, after which the distraction became negligible. Given the typical conversational rate of 110-150 words per minute \cite{speakingspeed}, even brief, 1-second feedback spans only about 2 to 3 words---well within the human short-term memory limit of approximately 3 to 5 chunks \cite{cowan2001magical}. This brief auditory feedback minimally disrupts conversational flow, allowing users to retain context and continue interacting naturally. Therefore, while the detailed information may induce initial distraction, users generally adjust and become accustomed to it if it is well designed.

The desire to improve speaking skills stands out as a key motivator, encouraging users to adapt to potential distractions. The stronger the user's goal to achieve meaningful outcomes, the higher their capacity to endure and benefit from the inherent distractions within the habit reversal process. P19 echoed this sentiment, expressing a keen interest in improving their speaking skills and reducing awkward and unwanted words. Consequently, they do not perceive the detailed information with auditory feedback as a distraction; instead, they appreciate it because it helps them improve quickly. Understanding these factors provides a foundation for tailoring methodologies to individual preferences, fostering a more personalized and effective habit-reversal experience.

\subsubsection{Trade-off: Ubiquitous Practice and User Endurance} As real-time auditory intervention systems become more integrated into daily life, maintaining user engagement without inducing fatigue is crucial. Ubiquitous intervention, while potentially effective, can risk user overload if not thoughtfully implemented. To mitigate this, adaptive scheduling can allow users to control the frequency and intensity of coaching sessions or specify contexts in which the coach should activate.

\subsection{Expanding Wearable Real-Time Auditory Interventions}

\subsubsection{Incorporating Contextual Awareness.} 
Dynamically adjusting feedback based on user context—such as emotional state \cite{hollis2018being} or gesture cues—may enhance both user experience and system effectiveness. In the case of delivering auditory feedback for unwanted words, it is important that the system consider not only what was said, but also how and when it was said. For example, signs of frustration or stress detected through vocal tone may indicate that the user is already cognitively burdened. In such cases, reducing the frequency or intensity of feedback could help prevent additional discomfort. Similarly, nonverbal signals such as posture, facial expression, or hesitation gestures may suggest agitation or uncertainty, signaling the system to delay or soften feedback delivery.

Moreover, not all unwanted words are inherently negative. Depending on the user’s context, filler words or informal expressions may serve communicative functions—such as emphasizing a point or maintaining conversational flow. Automatically flagging these without considering intent may disrupt the user unnecessarily and reduce the system’s utility. Understanding the contextual role of such expressions is therefore critical for effective intervention. This adaptability is particularly important in multilingual or multicultural settings, where rigid feedback may be misinterpreted. Context-aware strategies not only enhance relevance but also align with best practices in behavior change, which emphasize tailoring interventions to individual needs. Incorporating emotional and behavioral cues into feedback design offers a promising path toward more responsive and empathetic wearable assistants.

\subsubsection{Enhancing Multilingual and Holistic Behavioral Interventions.} 
This study focuses primarily on English, but the core functionality of the \WSCoach{} system can be extended to other languages. It detects unwanted words and delivers auditory feedback—a mechanism adaptable to any language. While the patterns of unwanted word usage may vary across languages, the underlying approach remains consistent. The system can be configured to identify language-specific unwanted expressions, ensuring that feedback effectively raises user awareness and supports behavior modification across diverse linguistic contexts.

Beyond unwanted speech, the system's core functionality can be applied to other behavioral habits. For example, it could support improvements in social behavior—such as prompting users to greet others—or help manage unconscious negative facial expressions by providing real-time auditory feedback that increases self-awareness. Making the system language-agnostic and expanding its scope to broader behavioral interventions would significantly enhance its applicability, establishing \WSCoach{} as a more versatile tool for supporting holistic personal development.

\section{Limitations and Future Work}

The present research calls for more work to address its limitations, both in terms of the study and the prototype.

\paragraph{System Limitations}
False detections occasionally occurred when bystanders (i.e., experimenters) used words from the user’s list of unwanted terms. In such cases, the system could not reliably differentiate the user’s voice from others, resulting in unintended feedback. This limitation stems from hardware constraints in most commercial smart glasses, including those used in our study, which lack spatial audio filtering or beamforming capabilities for speaker diarization. Consequently, the system could not directionally isolate the user’s speech. To address this, future iterations could incorporate multi-microphone spatial filtering to better isolate the user’s voice and reduce false positives.

\paragraph{Implementation Limitations}
The current implementation of \WSCoach{} runs on a GPU-enabled laptop connected to the smart glasses via Bluetooth, enabling high-accuracy, low-latency speech detection. While this setup supports reliable performance in typical indoor scenarios (e.g., indoor meetings or presentations with slides), it may limit portability and usability in mobile or outdoor contexts. Advances in on-device processing—such as edge AI accelerators in smartphones (e.g., Apple Neural Engine, Qualcomm Hexagon DSP)—could enable real-time inference directly on mobile or wearable devices. These developments would make it feasible to deploy \WSCoach{} in more mobile or resource-constrained environments without compromising performance.

\paragraph{Evaluation Limitations}
Our findings, though promising in a laboratory environment that mimicked typical indoor conditions, require further validation in longer-term and more diverse real-world contexts to enhance ecological validity. Although we minimized strict controls—allowing ambient variability akin to everyday settings—the generalizability of our results may still be constrained by limited environmental and demographic diversity. Expanding participant demographics and incorporating multilingual support are important next steps toward broadening the applicability of \WSCoach{}. We also acknowledge that adapting the system to handle auditory interference in more challenging environments, such as noisy outdoor settings, will be necessary for reliable deployment across varied contexts.

\section{Conclusion}

Our study sheds light on the potential of auditory feedback as a tool for enhancing self-awareness and reducing the use of unwanted words in daily conversations. The findings underscore the importance of selecting the appropriate type, duration, and delay time of feedback to achieve targeted speech corrections. These insights offer guidance to the development of speech intervention tools and the understanding of speech behavior correction. Looking forward, future work could focus on developing a general wearable real-time behavior coach and enhance people's awareness about their improper behavior in various application scenarios.

\begin{acks}
This research is supported by the National Research Foundation Singapore and DSO National Laboratories under the AI Singapore Programme (Award Number: AISG2-RP-2020-016). 
The CityU Start-up Grant (No. 9610677) and Guangxi Science and Technology Base and Talent Special Project (No. guikeAD23026230) also provide partial support.
Any opinions, findings, conclusions, or recommendations expressed in this material are those of the author(s) and do not reflect the views of the National Research Foundation, Singapore.
We extend our gratitude to all members of the Synteraction Lab for their help in completing this project. We also thank the reviewers for their valuable feedback.
\end{acks}

\bibliographystyle{ACM-Reference-Format}
\bibliography{Paper/reference}

\appendix
\newpage{}
\section{Methodological Transparency \& Reproducibility Appendix (META)}
\label{appendix:meta}

This section describes the supplementary material we shared in the Open Science Framework repository (OSF) at\\ \url{https://osf.io/6vhwn/?view_only=489498d3ac2d4703a17475fc6ca65dfa} (anonymous link). The OSF project is currently private and anonymized for blind reviewing, and will become public once the paper is accepted.

\subsection{Preregistration}

The preregistration of our evaluation experiment (Sec~\ref{sec:eval}) is available at\\ \url{https://osf.io/yp5xq/?view_only=a3e22795552a4030a95ab8ed02c82a8e} (anonymous link). 

\subsection{Experimental Data}

The data for our evaluation experiment is available in our OSF project as a spreadsheet with three tabs:

\begin{itemize}[leftmargin=*]
\item \textit{Frequency}. This table contains the normalized frequency of unwanted words for each participant (rows 1 to 24) and each session (columns S1 to S8). The column System specifies the experimental group to which each participant was randomly assigned.
\item \textit{Original Data}. This table contains the raw data from which the normalized frequencies were calculated. Again, rows are participants 1 to 24, and for each participant $\times$ session combination, the table reports the participant's total speaking time, and the full length of the conversation. The rows are further broken down into individual unwanted words (those identified by the participant in the first session). For each combination of participant $\times$ unwanted word $\times$ session, the table reports the number of times the unwanted word was uttered and the corresponding frequency (words per minute). Finally, for each participant $\times$ session, the table reports the normalized frequency for all unwanted words confounded (column Average).
\item \textit{Questionnaire}. This table contains participant responses to the six subjective questions listed in \autoref{tab:Study2Measurements} and their score to the Nasa-TLX questionnaire.
\end{itemize}

\subsection{Interview Materials}
In our OSF project, we share the interview questions used in the training and post-training phases.

\subsection{\WSCoach{} Software}

In our OSF project, we share the python source code for the WSCoach software we used in our evaluation. We ran this software on a desktop computer with a 3080Ti Graph Card connected to a Huawei Eyewear via Bluetooth, but any audio mic (e.g., Bluetooth earbuds) can be used as the audio source and laptops with graphic cards can run the program. The computer should have Python3 installed. Further instructions are available in the README file.

\subsection{Additional Information About \Orai{}}
\label{appendix:orai_app}

\begin{figure}[hptb]
	\centering
	\includegraphics[width=0.75\linewidth]{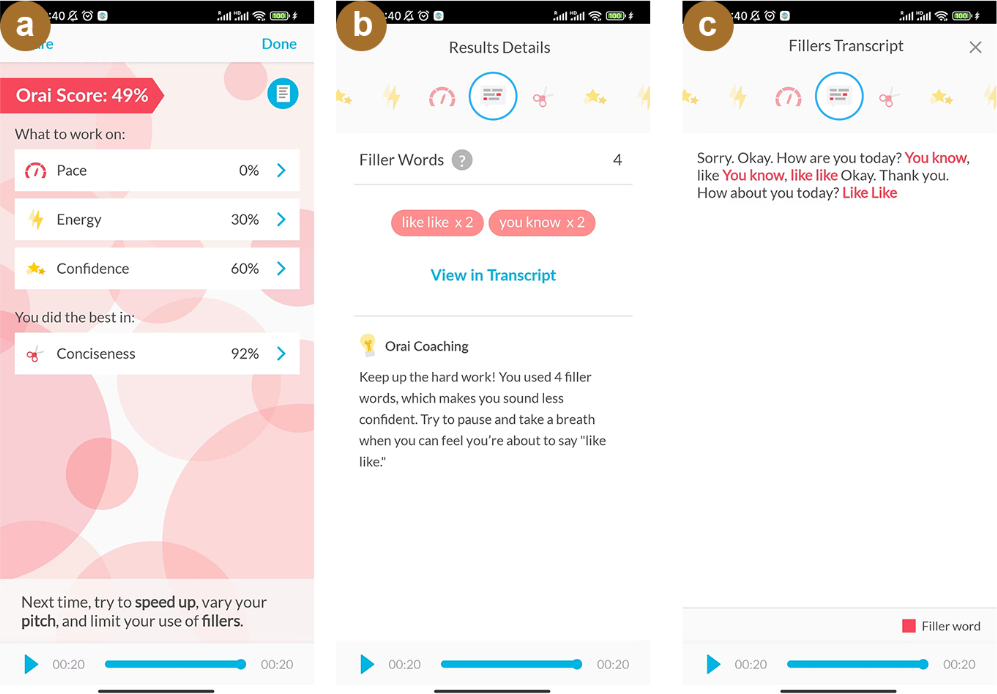}
	\caption{The \Orai{} application. (a) \Orai{} provides feedback on key communication metrics. (b) \Orai{} Lists the count of filler words or unwanted words after conversation visually. (c) Users can see the highlighted filler words in audio transcription and play the recording.}
	\label{fig:orai_app}
\end{figure}

Figure~\ref{fig:orai_app} (also available in the OSF project) shows screenshots of the \Orai{} application's user interface, enabling diverse post-conversation analyses. For more information please visit \url{https://orai.com/}.

\subsection{Full Analysis of Secondary Measurements}

Here we report the full analysis of secondary measurements, which we only partially reported in Sec~\ref{sec:secondary-measurements} for space reasons.

\begin{itemize}[leftmargin=*]
\item \textit{Confidence in Speaking.} The mean response for this question was 4.8, 95\% CI [4.0, 5.5] for \WSCoach{} and 4.5, 95\% CI [3.6, 5.4] for \Orai{}, see \autoref{fig:Long phase} (a). The $U$-statistic for the difference is $U = 71$, $p=.99$. Therefore, we have no evidence that \WSCoach{} and \Orai{} differed in terms of Confidence in Speaking.

\item \textit{Awareness of Speaking Unwanted Words.} The mean score was 5.8, 95\% CI [5.2, 6.4] for \WSCoach{} and 5.5, 95\% CI [5.2, 5.8] for \Orai{}, see \autoref{fig:Long phase} (b). The $U$-statistic for the difference is $U = 54$, $p=.27$. Therefore, we have no evidence that \WSCoach{} and \Orai{} differed in terms of Awareness of Speaking Unwanted Words.

\item \textit{Self-Ranked Conversation Quality.} The mean score was 5.4, 95\% CI [4.9, 6.0] for \WSCoach{} and 5.1, 95\% CI [4.5, 5.7] for \Orai{}, $U = 54$, $p=.31$, see \autoref{fig:Long phase} (c). Again we have no evidence for a difference on this metric.

\item \textit{Distraction of Feedback.} The mean score was 4.1, 95\% CI  [3.0, 5.2] for \WSCoach{} and 1.8, 95\% CI [1.1, 2.4] for \Orai{}, $U = 117$, $p=.003$, see \autoref{fig:Long phase} (d). Therefore, we have good evidence that \WSCoach{} generated more distraction on average than \Orai{} during conversations.
    
\item \textit{NASA TLX for Being Aware of Unwanted Words.} The mean score was 41, 95\% CI [25, 57] for \WSCoach{} and 30.64, 95\% CI  [20, 41] for \Orai{}, $U = 55$, $p=.35$, see \autoref{fig:Long phase} (e). Again, we have no evidence of a difference here either.

\item \textit{Confidence in Conversation (post-training phase).} The mean score was 5.3, 95\% CI  [4.9, 5.6] for \WSCoach{} and 4.3, 95\% CI [3.8, 4.9] for \Orai{}, $U = 28$, $p=.01$, see \autoref{fig:Long phase} (f). Therefore, we have evidence that \WSCoach{} made people feel more confident in their conversation than \Orai{} after the training was over. 

\item \textit{Awareness of Avoiding Unwanted Words (post-training phase).} The mean response was 5.4, 95\% CI [4.9, 5.9] for \WSCoach{} and 4.8, 95\% CI [4.0, 5.6] for \Orai{}, see \autoref{fig:Long phase} (g). The $U$-statistic is $U = 53$, $p=.25$. Therefore, we have no evidence that \WSCoach{} and \Orai{} differed in terms of Awareness of Avoiding Unwanted Words.

\end{itemize}

\subsection{Auditory feedback audio samples}

Our OSF project contains audio samples we used in our pilot studies. There is one example for each of the five feedback techniques we used: Earcon, Spearcon, Spindex, Speech, and Lyricon. We also provide an example of Auditory Icon.

\end{document}